\documentclass{emulateapj}
\usepackage{apjfonts}
\usepackage{graphicx}
\usepackage{xspace}
\usepackage{graphicx}
\usepackage{longtable}
\usepackage{psfig}

\newcommand{\Msun}      {\mbox{$\rm\,M_{\mathord\odot}$}}

\begin{document}

\lefthead{Galactic {\em NuSTAR} Serendipitous Sources}
\righthead{Tomsick et al.}

\submitted{Accepted by ApJ}

\def\lsim{\mathrel{\lower .85ex\hbox{\rlap{$\sim$}\raise
.95ex\hbox{$<$} }}}
\def\gsim{\mathrel{\lower .80ex\hbox{\rlap{$\sim$}\raise
.90ex\hbox{$>$} }}}

\title{Galactic Sources Detected in the {\em NuSTAR} Serendipitous Survey}

\author{John A. Tomsick\altaffilmark{1}, 
George B. Lansbury\altaffilmark{2},
Farid Rahoui\altaffilmark{3}, 
Ma\"ica Clavel\altaffilmark{1},
Francesca M. Fornasini\altaffilmark{4}, 
JaeSub Hong\altaffilmark{4}, 
James Aird\altaffilmark{2},
David M. Alexander\altaffilmark{5},
Arash Bodaghee\altaffilmark{6},
Jeng-Lun Chiu\altaffilmark{1},
Jonathan E. Grindlay\altaffilmark{4},
Charles J. Hailey\altaffilmark{7},
Fiona A. Harrison\altaffilmark{8},
Roman A. Krivonos\altaffilmark{9},
Kaya Mori\altaffilmark{7}, 
Daniel Stern\altaffilmark{10}}

\altaffiltext{1}{Space Sciences Laboratory, 7 Gauss Way, University of California, 
Berkeley, CA 94720-7450, USA}

\altaffiltext{2}{Institute of Astronomy, University of Cambridge, Madingley Road, 
Cambridge CB3 0HA, UK}

\altaffiltext{3}{Department of Astronomy, Harvard University, 60 Garden Street, 
Cambridge, MA 02138, USA}

\altaffiltext{4}{Harvard-Smithsonian Center for Astrophysics, 60 Garden Street,
Cambridge, MA 02138, USA}

\altaffiltext{5}{Centre for Extragalactic Astronomy, Department of Physics, University 
of Durham, South Road, Durham DH1 3LE, UK}

\altaffiltext{6}{Georgia College and State University, Milledgeville, GA 31061, USA}

\altaffiltext{7}{Columbia Astrophysics Laboratory, Columbia University, New York, NY 
10027, USA}

\altaffiltext{8}{California Institute of Technology, 1200 East California Boulevard, 
Pasadena, CA 91125, USA}

\altaffiltext{9}{Space Research Institute of the Russian Academy of Sciences, 
Profsoyuznaya Str. 84/32, 117997, Moscow, Russian}

\altaffiltext{10}{Jet Propulsion Laboratory, California Institute of Technology, 
4800 Oak Grove Drive, Pasadena, CA 91109, USA}

\begin{abstract}

The {\em Nuclear Spectroscopic Telescope Array (NuSTAR)} provides an improvement in sensitivity at energies above 10\,keV by two orders of magnitude over non-focusing satellites, making it possible to probe deeper into the Galaxy and Universe.  Lansbury and collaborators recently completed a catalog of 497 sources serendipitously detected in the 3--24\,keV band using 13\,deg$^{2}$ of {\em NuSTAR} coverage.  Here, we report on an optical and X-ray study of 16 Galactic sources in the catalog.  We identify eight of them as stars (but some or all could have binary companions), and use information from {\em Gaia} to report distances and X-ray luminosities for three of them.  There are four CVs or CV candidates, and we argue that NuSTAR~J233426--2343.9 is a relatively strong CV candidate based partly on an X-ray spectrum from {\em XMM-Newton}.  NuSTAR~J092418--3142.2, which is the brightest serendipitous source in the Lansbury catalog, and NuSTAR~J073959--3147.8 are LMXB candidates, but it is also possible that these two sources are CVs.  One of the sources is a known HMXB, and NuSTAR~J105008--5958.8 is a new HMXB candidate, which has strong Balmer emission lines in its optical spectrum and a hard X-ray spectrum.  We discuss the implications of finding these HMXBs for the surface density (log$N$-log$S$) and luminosity function of Galactic HMXBs.  We conclude that, with the large fraction of unclassified sources in the Galactic plane detected by {\em NuSTAR} in the 8--24\,keV band, there could be a significant population of low luminosity HMXBs.

\end{abstract}


\section{Introduction}

The {\em NuSTAR} serendipitous source survey is a systematic analysis of all 
{\em NuSTAR} observations excluding the core Galactic Survey programs, the 
Galactic center and the Norma regions \citep{hong16,fornasini17}, and excluding
the dedicated Extragalactic Survey programs: COSMOS, ECDFS, EGS, GOODS-N, and UDS
\citep[e.g.,][]{civano15,mullaney15}.  After detecting the sources with {\em NuSTAR}
and looking for counterparts at other wavelengths, we have been performing
ground-based optical spectroscopy to identify the sources.  This program is
described in \cite{alexander13} and \cite{aird15}, and the full 40-month
catalog is published in \cite{lansbury17}.

The {\em NuSTAR} serendipitous survey takes advantage of the sensitivity of
{\em NuSTAR} in the hard X-ray band \citep{harrison13}.  Although the bandpass
of {\em NuSTAR} is 3--79\,keV, the analysis for the serendipitous survey has
been carried out in the 3--24\,keV band where the source-to-background ratio
is higher for most source types.  Thus, in energy, the {\em NuSTAR}
survey is intermediate between surveys and catalogs constructed in the soft
X-ray band with {\em ASCA} \citep[0.7--10\,keV,][]{sugizaki01}, {\em XMM-Newton}
\citep[0.2--12\,keV,][]{rosen16}, and the {\em Chandra X-ray Observatory}
\citep[0.5--7\,keV,][]{evans10} and at higher energies with 
{\em Swift}/BAT \citep[15--55\,keV,][]{ajello12} and with
{\em INTEGRAL} \citep[17--100\,keV,][]{bird16}.  In terms of coverage,
BAT and {\em INTEGRAL} have large fields of view (FOVs), and they have
observed the entire sky.  The {\em Chandra} FOV is comparable to
{\em NuSTAR}'s, while {\em XMM}'s is larger (0.2\,deg$^{2}$ for {\em XMM}
vs. 0.04\,deg$^{2}$ for {\em NuSTAR}).  Coupled with the fact that
{\em Chandra} and {\em XMM} have been observing for much longer than
{\em NuSTAR}, their coverage is larger (approaching 1000\,deg$^{2}$ for
{\em XMM} compared to 13\,deg$^{2}$ for {\em NuSTAR} in the 40-month catalog).
However, {\em NuSTAR}'s coverage will grow over time, and it is providing
the first sensitive survey in the 8--24\,keV band.  Its sensitivity is
approximately two orders of magnitude better than that of BAT and
{\em INTEGRAL}, pushing into new discovery space \citep{lansbury17}.

The source types that {\em INTEGRAL} and {\em Swift}/BAT have detected
in the largest numbers \citep{bird16,va10,krivonos12} are Active Galactic
Nuclei (AGN), Low-Mass X-ray Binaries (LMXBs), High-Mass X-ray Binaries
(HMXBs), and Cataclysmic Variables (CVs).  While these accreting black
holes, neutron stars, and white dwarfs are the most common types,
{\em INTEGRAL} also detects significant numbers of non-accreting compact
objects, including pulsar wind nebulae and magnetars.  While \cite{lansbury17}
reported basic information (e.g., positions, count rates, and fluxes with
{\em NuSTAR}) for all of the serendipitously detected sources, the
scientific focus of that work is the extragalactic sources, including
the identification of a large group of hard X-ray selected AGN.  The
{\em NuSTAR} AGN have a median redshift of $z = 0.56$, which is about
an order of magnitude higher than the median redshift of the brighter
AGN in the BAT sample.  In the current work, we are reporting on the
Galactic sources, and we can expect the following advances:  1.~for all
Galactic source types, the X-ray luminosities are such that the search
volumes for existing surveys do not extend to the other side of the Galaxy,
so going deeper increases the search volume and essentially guarantees that
new hard X-ray sources will be found; 2.~considering HMXBs, \cite{lutovinov13}
show that {\em INTEGRAL}'s search volume for sources with luminosities
$>$$10^{35}$\,erg\,s$^{-1}$ extends only a small distance past the center
of the Galaxy; and 3.~we are also searching for closer sources
at lower luminosity (e.g., HMXBs with weak stellar winds), and if we do
not find such sources, this will have implications for the luminosity
functions of the different source types.

The known population of HMXBs has increased in size by nearly a factor of
three over the past decade due to coverage of the Galactic plane at $>$17\,keV
provided by {\em INTEGRAL} \citep{walter15}.  Even though {\em INTEGRAL} has
given us a much better estimate of the total number of HMXBs in the Galaxy, we
could still be missing a large part of the faint population, which is important
for determining if the luminosity function for persistent HMXBs breaks below
$\sim$$10^{35}$\,erg\,s$^{-1}$ \citep{lutovinov13}.  One reason why this question
is interesting is that HMXBs are the progenitors of compact object merger events,
and, as more of these are detected via gravitational waves \citep{abbott16}, we
anticipate obtaining a much more complete picture of HMXB evolution since a
significant fractions of HMXBs will evolve to NS-NS, NS-BH, or BH-BH binaries.
Constraining HMXB evolution may also provide information about the distant and
early Universe.  As HMXBs form and remain luminous for $\sim$10--30\,Myr after
a starburst, their X-ray emission can be used to trace the star formation rate in
distant galaxies \citep[e.g.,][]{mineo12}.  Also, it is possible that HMXBs
played a role in the heating and reionization of the early Universe
\citep[e.g.,][]{brorby16}.  While the high-luminosity end of the distribution
has the greatest impact, different phases of HMXB evolution will produce different
levels of X-ray emission; thus, knowing the total number of HMXBs, how many harbor
black holes as opposed to neutron stars, and how they evolve helps to constrain
the heating that they could have caused.

While the {\em NuSTAR} survey is much deeper than previous hard X-ray
surveys, the {\em XMM} and {\em Chandra} surveys extend to much lower
luminosities in the soft X-rays.  However, so many sources are found that
a very small fraction is followed up in any way.  Thus, {\em NuSTAR} also
plays the role of selecting sources that are already in the {\em XMM} or
{\em Chandra} catalogs.  While soft X-rays are produced by many types of
sources due to thermal processes, extending into the hard X-rays greatly
increases the fraction of sources with extreme physics (e.g., particle
acceleration, accretion shocks, relativistic jets, highly magnetic neutron
stars, and the strong gravity around black holes and neutron stars).  In
addition, sources in the Galactic plane can be obscured by interstellar
material or material local to the source, and observing above 8\,keV
decreases this bias.

In this paper, we describe how we defined our sample of 16 Galactic sources
from \cite{lansbury17} in Section 2.  Section 3 details the soft X-ray and
optical counterpart identifications found by \cite{lansbury17}, and we also
perform new searches of the SIMBAD database to determine if the nature of any
of the sources is known.  In Section 4, we report on the hard X-ray fluxes of
the sources, including a new {\em NuSTAR} measurement of NuSTAR~J092418--3142.2,
which is the brightest serendipitous source discovered.  For the nine sources
that have not already been identified in SIMBAD, we analyze their optical
spectra in Section 5 and use them to discuss their identifications.  In
Section 6, we analyze the {\em NuSTAR}, {\em XMM}, and {\em Swift} spectra
for the four sources that show optical emission lines, which may be a sign
that they have accretion disks.  In Section 7, we discuss the results, including
a detailed look at implications for the HMXB population in the Galaxy and
also a discussion of the incompleteness of the survey, especially regarding the
number of sources with optical spectroscopic identifications near the
Galactic plane.  Finally, in Section 8, we describe our conclusions and
discuss possibilities for future work.

\begin{figure*}
\plotone{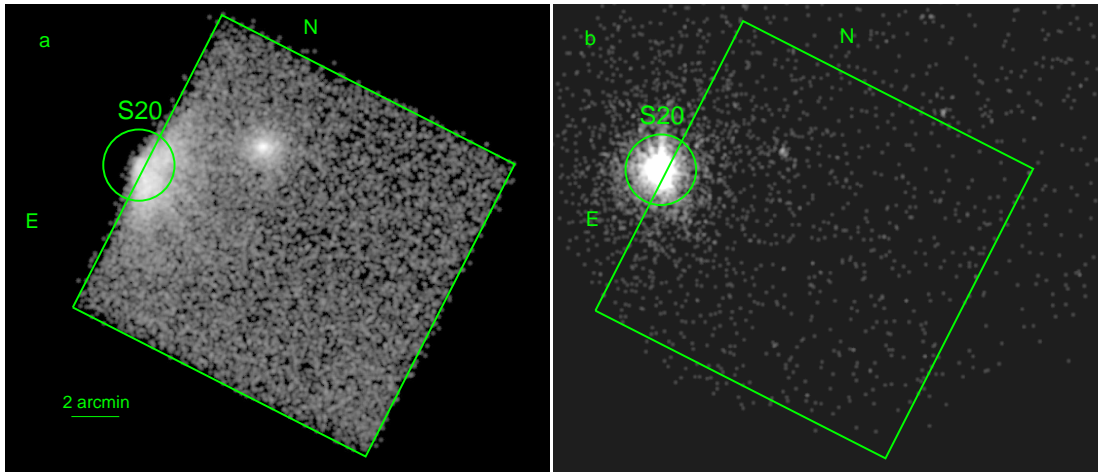}
\caption{\footnotesize {\em (a)} 3--79\,keV {\em NuSTAR} image for Focal Plane Module A
from ObsID 60061339002.  The primary target of the observation was 2MASX~J09235371--3141305.
The emission from the serendipitous source S20 (NuSTAR~J092418--3142.2) is labeled. {\em (b)}
A 0.5--10\,keV {\em Swift}/XRT image (from ObsID 00080674001) of the same field taken on 
2014 April 19 during the {\em NuSTAR} observation.  The XRT image provides full coverage of 
S20, and we use this observation to determine the position of NuSTAR~J092418--3142.2.
\label{fig:images_s20}}
\end{figure*}

\section{The Galactic {\em NuSTAR} Serendipitous Sources}

The primary \cite{lansbury17} 40-month catalog includes 497 detected
serendipitous sources (or ``serendips'') using $\sim$20\,Ms of {\em NuSTAR}
exposure with 13\,deg$^{2}$ of coverage.  From ground-based follow-up and
archival searches, classifications were obtained for 276 of the serendips
in the primary catalog, with 94\% of these (260) being AGN.  The remaining
16 sources are classified as Galactic sources based on having optical emission
or absorption lines at zero redshift.  \cite{lansbury17} also provide a
secondary catalog with 64 serendips found using a different source detection
approach from that used for the primary catalog, and five of these are
classified as Galactic sources.  Among the 21 sources classified as Galactic
in the primary and secondary catalogs, \cite{lansbury17} note that there is
uncertainty about the optical counterpart in five of these cases (for
NuSTAR~J080421+0504.9, NuSTAR~J102318+0036.5, NuSTAR~J202339+3347.7,
NuSTAR~J202351+3354.3, and NuSTAR~J202420+3347.7), and we save these for
future work (e.g., after more accurate X-ray positions are obtained, allowing
for optical or near-IR spectroscopy).

This study therefore focuses on 16 sources: 11 from the primary catalog and
five from the secondary catalog.  Table~\ref{tab:list} provides basic
information about the 16 sources, including the ID number from the
\cite{lansbury17} catalogs (the IDs starting with ``P'' and ``S'' are from the
primary and secondary catalogs, respectively), the source names, the positions
in Equatorial and Galactic coordinates, and the {\em NuSTAR} exposure time used
for the serendipitous survey.  With one exception (S20), the Equatorial
positions come directly from \cite{lansbury17}, and they have 90\% confidence
uncertainties of 14--22$^{\prime\prime}$, depending on the source detection
significance.  We note that the sources are widely spread in Galactic latitude,
from $b$ = --72.3$^{\circ}$ to $53.5^{\circ}$, and only six of the 16 sources
are within $10^{\circ}$ of the plane.  This does not match any single known
population of Galactic sources, suggesting that this first group of serendips
identified as being Galactic likely contains a significant number of nearby
sources with high latitude, and those at $\left| b\right| > 10^{\circ}$ are
dominated by active stars and CVs \citep{sazonov06}.  However, the serendips
in the Galactic plane are under-represented in the current study because they
are more difficult to classify.  As discussed in Sections 7.2 and 7.3, a large
fraction of the sources in the Galactic plane are currently unclassified due to
the challenges caused by crowding and extinction.

For S20, the source was at the edge of the {\em NuSTAR} field of view for the 
observation in which the source was discovered, and the 3--79\,keV image 
is shown in Figure~\ref{fig:images_s20}a.  While only a very rough position for
S20 can be obtained from the {\em NuSTAR} data, a {\em Swift} X-ray Telescope
(XRT) observation occurred on 2014 April 19 during the {\em NuSTAR} observation.
The XRT field of view is somewhat larger, and the S20 point spread function
is fully covered (Figure~\ref{fig:images_s20}b).  We used {\ttfamily xrtcentroid}
to constrain the position of S20 to be R.A. (J2000) = $09^{\rm h}24^{\rm m}18^{\rm s}.17$, 
Decl. (J2000) = --$31^{\circ}42^{\prime}17.\!^{\prime\prime}2$ with an 90\% confidence
uncertainty of $3.5^{\prime\prime}$, and this is the position that is given in 
Table~\ref{tab:list}.

\begin{figure}
\hspace{-0.5cm}
\epsscale{1.2}
\plotone{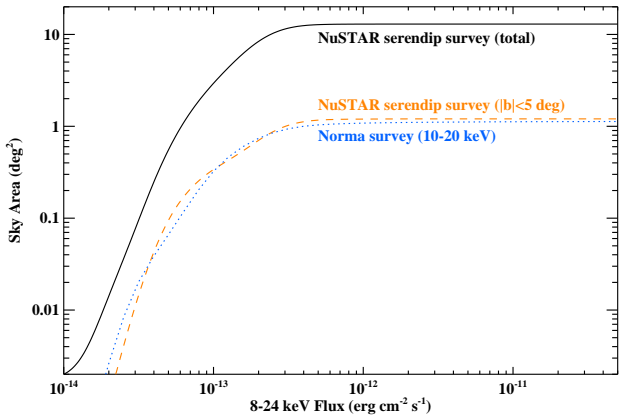}
\caption{\footnotesize Sky coverage curves for {\em NuSTAR} surveys.  The 
black curve shows the 8--24\,keV curve for the full serendipitous source
survey, which plateaus at 13\,deg$^{2}$ \citep{lansbury17}.  The orange 
dashed line is also for the serendipitous source survey, but it only includes
the coverage at Galactic latitudes between --5$^{\circ}$ and $5^{\circ}$, and
plateaus at 1.2\,deg$^{2}$.  The blue dotted line is a 10--20\,keV curve for 
the Norma spiral arm region survey \citep{fornasini17}.\label{fig:coverage}}
\end{figure}

While the full 13\,deg$^{2}$ of sky coverage is for all Galactic latitudes and
sources fluxes down to $\sim$$4\times 10^{-13}$\,erg\,cm$^{-2}$\,s$^{-1}$ (8--24\,keV), 
Figure~\ref{fig:coverage} shows that there is still significant coverage at
fluxes that are an order of magnitude lower.  While stars and CVs are somewhat
more concentrated toward the Galactic plane, their space densities are high enough
that they can be found at any Galactic latitude \citep{sazonov06,revnivtsev08,pk12}.
On the other hand, HMXBs have a comparatively low space density and are strongly
concentrated toward the Galactic plane.  In the \cite{bird16} {\em INTEGRAL} catalog,
after removing HMXBs in the Magellanic Clouds, 104 of 105 HMXBs have Galactic
latitudes between --$4.1^{\circ}$ and $5.2^{\circ}$; thus, in Figure~\ref{fig:coverage},
we also show the sky coverage for the serendipitous source survey for observations 
within $5^{\circ}$ of the Galactic plane.  Although this is a large reduction 
from 13\,deg$^{2}$ to 1.2\,deg$^{2}$, the survey has very similar coverage to
the Norma spiral arm survey\footnote{We note that the Norma study used a slightly
different hard X-ray energy band of 10--20\,keV.} \citep{fornasini17}, which is one 
of the primary {\em NuSTAR} Galactic plane surveys.

\section{X-ray/Optical Counterparts and SIMBAD Identifications}

Table~\ref{tab:list_soft} lists the soft X-ray counterparts identified by
\cite{lansbury17} for each of the {\em NuSTAR} serendips.  They come from
catalogs or analysis of archival observations with {\em XMM-Newton},
{\em Chandra}, and {\em Swift}.  While the soft X-ray position uncertainties
depend on a variety of factors (e.g., number of source counts detected, satellite
point spread function, off-axis angle, and systematic offsets), they are typically
near $1^{\prime\prime}$ for {\em Chandra}, a few arcseconds for {\em XMM-Newton},
and several arcseconds for {\em Swift}.  The positions of the soft X-ray sources
are provided along with the separation between the positions reported in
Table~\ref{tab:list} and the soft X-ray position.  For the 15 sources with
{\em NuSTAR} positions, the separations range from $2.8^{\prime\prime}$ to
$26.5^{\prime\prime}$, with two out of 15 being slighly beyond the 90\% confidence 
{\em NuSTAR} error circle, which is consistent with the expected statistical
distribution of separations.  For S20, there is a {\em Chandra} source,
CXO~J092418.2--314217, within $0.7^{\prime\prime}$ of the {\em Swift} position.

The optical counterparts are listed in Table~\ref{tab:list_opt} with the
optical catalog where the source is found, the Equatorial coordinates of
the counterparts, the separation between the soft X-ray and optical
positions, and the $R$-band magnitude.  For the four {\em Chandra}
sources (P146, S20, S37, and P408), the separations are $0.15^{\prime\prime}$,
$0.24^{\prime\prime}$, $0.56^{\prime\prime}$, and $0.26^{\prime\prime}$, respectively,
which is consistent with {\em Chandra}'s sub-arcsecond accuracy.  For
the ten {\em XMM-Newton} sources, the average separation is $1.7^{\prime\prime}$,
with a range between $0.50^{\prime\prime}$ (for P82) and $4.25^{\prime\prime}$
(for P316).  The separations are reasonable considering the accuracy of the
{\em XMM-Newton} positions, but the identifications for sources with the largest
separations (P316 and perhaps P376) are worth confirming with future
observations with {\em Chandra} to improve the X-ray position constraints.
The two {\em Swift} source separations are $3.67^{\prime\prime}$ for P98 and
$1.87^{\prime\prime}$ for P497, which are consistent with the {\em Swift} position
uncertainties.

For each optical position, we searched the SIMBAD astronomical database
\citep{wenger00} to determine if the sources have already been classified,
and the results of the searches are summarized in Table~\ref{tab:list_simbad}.
We found SIMBAD matches in nine of 16 cases, and the source type is known in
seven cases.  Often, sources have several different names (or ``identifiers''),
and, in Table~\ref{tab:list_simbad}, we give one of the identifiers as well
as the number of identifiers.  We also indicate the wavelength in
which the sources have been previously detected.  For serendips
S1, P98, and P146, this is the first time that these sources have been reported
as X-ray sources.  Of the sources that have been classified, there are
four bright stars:  S1 is HD 1165 with $R = 8.16$; P146 is TYC 7654-3811-1
with $R = 9.55$; S37 is HD 109573B with $R = 11.8$; and P340 is TYC 3866-132-1
with $R = 13.98$.  In three cases, we obtained the classifications before
performing optical follow-up, but, for P146, the optical spectrum we obtained
is shown in \cite{lansbury17}, confirming its stellar nature.  There are also
two Cataclysmic Variables (CVs):  Serendip P82 is RX~J0425.6--5714, which is a
polar-type magnetic CV \citep{halpern98}; and serendip P98 is the nova-like
CV V1193~Ori with properties similar to a non-magnetic CV in outburst
\citep{bond87}.  We obtained optical spectra for both of these, and the strong
Balmer series emission lines and optical continuum with low extinction
\citep{lansbury17} are consistent with the sources being CVs.  Finally,
serendip S43 is a known HMXB, 2RXP~J130159.6--635806.  It is an accreting pulsar
with a spin period near 700\,s, and a full analysis of the {\em NuSTAR} data
is reported in \cite{krivonos15}.

Two of the other Galactic serendips have matches in SIMBAD, but they are not
classified.  Serendip P194 is known to be an X-ray source, 2XMM J095839.2+690910,
but its nature is uncertain.  Also, there is an apparent match between
serendip P497 and the {\em ROSAT} source 1RXS J233427.8--234419.  For these
two sources and the other seven that do not have any matches in SIMBAD, we
consider possible classifications in Section 5.

\section{The Hard X-ray Emission from the Galactic Serendips}

\begin{figure}
\hspace{-0.5cm}
\epsscale{1.2}
\plotone{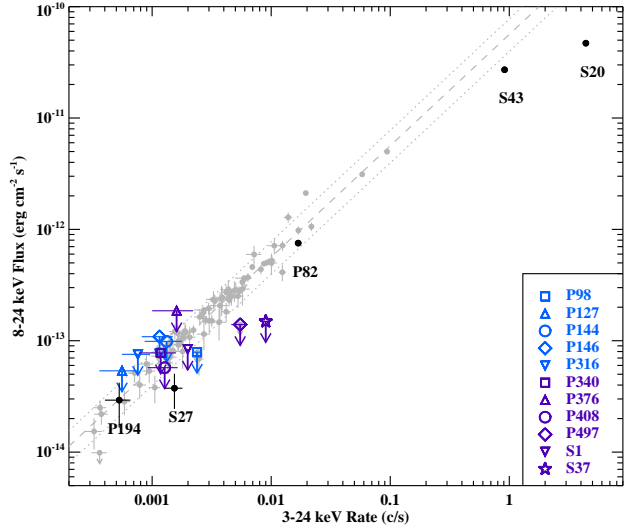}
\caption{\footnotesize The 8--24\,keV flux vs. the 3--24\,keV {\em NuSTAR} count rate.  The 
values are taken from \cite{lansbury17}, except for S20 (see text).  The gray points are AGN 
detected at very high significance in the primary serendipitous source catalog.  The dashed 
gray line corresponds to a power-law spectrum with a photon index of $\Gamma = 1.8$, and the
dotted gray lines are for $\Gamma = 1.3$ and $\Gamma = 2.3$.  The 
black points mark Galactic sources detected at 8--24\,keV, and the blue and purple points 
mark Galactic sources that are not detected in the 8--24\,keV bandpass, and we show the
upper limits.\label{fig:hf}}
\end{figure}

\cite{lansbury17} includes {\em NuSTAR} count rates and fluxes for the serendips
in the primary and secondary catalogs.  They are reported for the 3--24\,keV,
3--8\,keV, and 8--24\,keV bands.  In Figure~\ref{fig:hf}, we plot the 8--24\,keV 
flux vs. the 3--24\,keV rate.  In addition to the 16 Galactic serendips, we plot
the same quantities for the 81 AGN in the primary catalog with a false probability
of detection below $10^{-20}$.  A typical power-law photon index for AGN is
$\Gamma = 1.8$, and the flux/rate relationship for such a spectrum is indicated
as a dashed line on Figure~\ref{fig:hf}.  Thus, the location of the Galactic
sources in the diagram allows for a comparison to the hardness of the AGN spectra.

Of the seven sources with SIMBAD classifications, the four bright stars and the
non-magnetic CV are not detected at 8--24\,keV (see Figure~\ref{fig:hf}).  S37
is significantly softer than the AGN, the sources P98 and S1 are at least moderately 
softer, and the limits on the other two (P146 and P340) are not constraining.  The
magnetic CV (P82) has a hard spectrum, which may be slightly surprising since P82
is a polar-type CV, and intermediate polars typically have harder spectra than polars
\citep{revnivtsev08}.  Section 7.1 provides more details on the properties of the 
previously known CVs (P82 and P98).  The HMXB (S43) is close to the hardness of 
the AGN, but it is somewhat below the $\Gamma = 1.8$ line. This is not very surprising 
because, although accreting pulsars in HMXBs (like S43) typically have very hard 
spectra below 10\,keV, their spectra have cutoffs starting near 10\,keV.  In fact, 
the S43 spectrum has $\Gamma\sim 1.4$, but its exponential cutoff starts at 
$\sim$7\,keV \citep{krivonos15}.

Five of the 16 Galactic serendips are detected in the 8--24\,keV band.  In
addition to P82 and S43, the sources S20, S27, and P194 are detected.  The hardnesses 
for S27 and P194 put them close to the $\Gamma = 1.8$ line, but S20, which is the 
brightest {\em NuSTAR} serendip detected to date, is softer.  Given that S20 does 
not have any absorption or emission lines in its optical spectrum (see Section~5), 
the fact that its X-ray spectrum is too soft for it to be an AGN is an important
confirmation that it is Galactic.  As described in Section~2, S20 was mostly off
the active area of the detector when it was discovered, and \cite{lansbury17} do not
provide count rates and fluxes for S20.  However, on 2016 December 10, we obtained
a dedicated observation of S20 (= NuSTAR J092418--3142.2 = CXO J092418.2--314217)
with {\em NuSTAR} (ObsID 30201014002) and {\em XMM-Newton} (ObsID 0790620101).
While we plan to report on the details of the observation in a future paper, we
used the {\em NuSTAR} data from ObsID 30201014002 to determine the values for
S20 shown in Figure~\ref{fig:hf}.

\section{Optical Spectroscopy and Possible Source Classifications}

\begin{figure*}
\centering
\begin{minipage}[l]{0.47\textwidth}
\includegraphics[width=\textwidth]{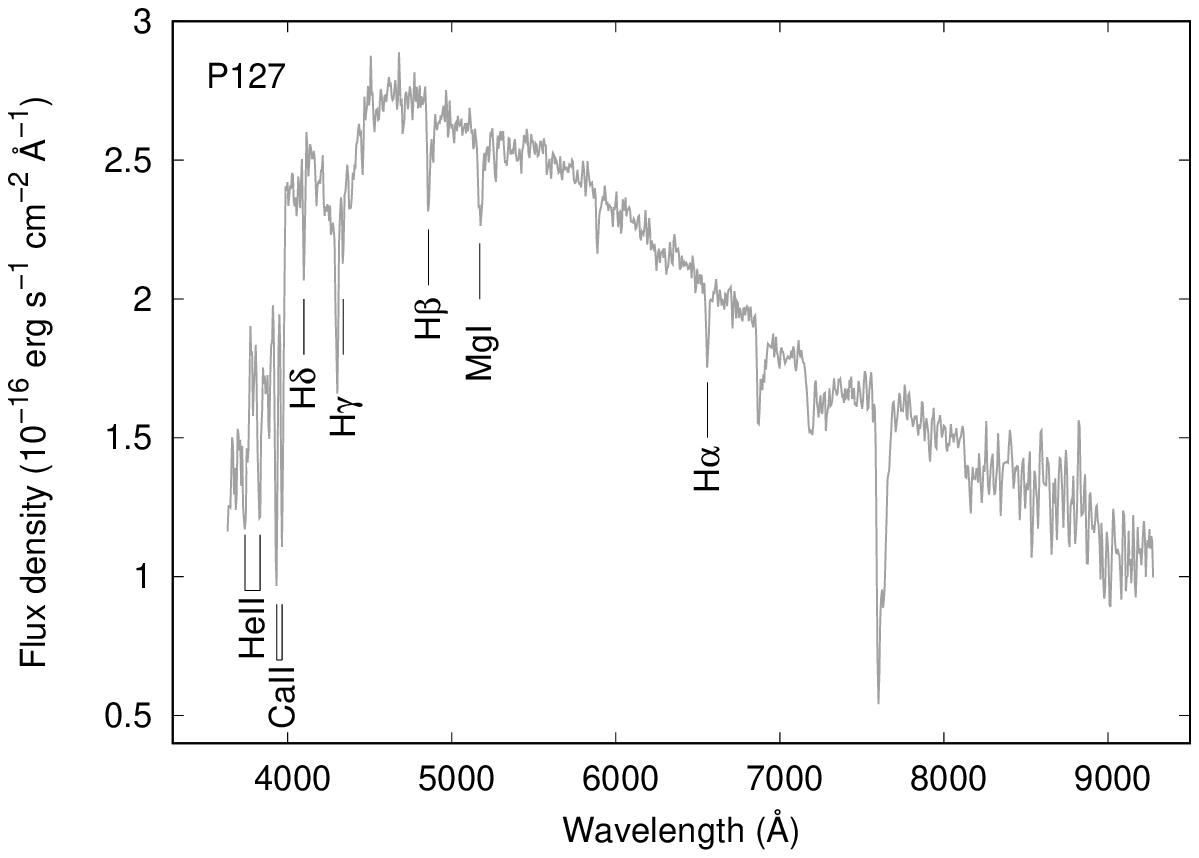}
\end{minipage}
\begin{minipage}[l]{0.47\textwidth}
\includegraphics[width=\textwidth]{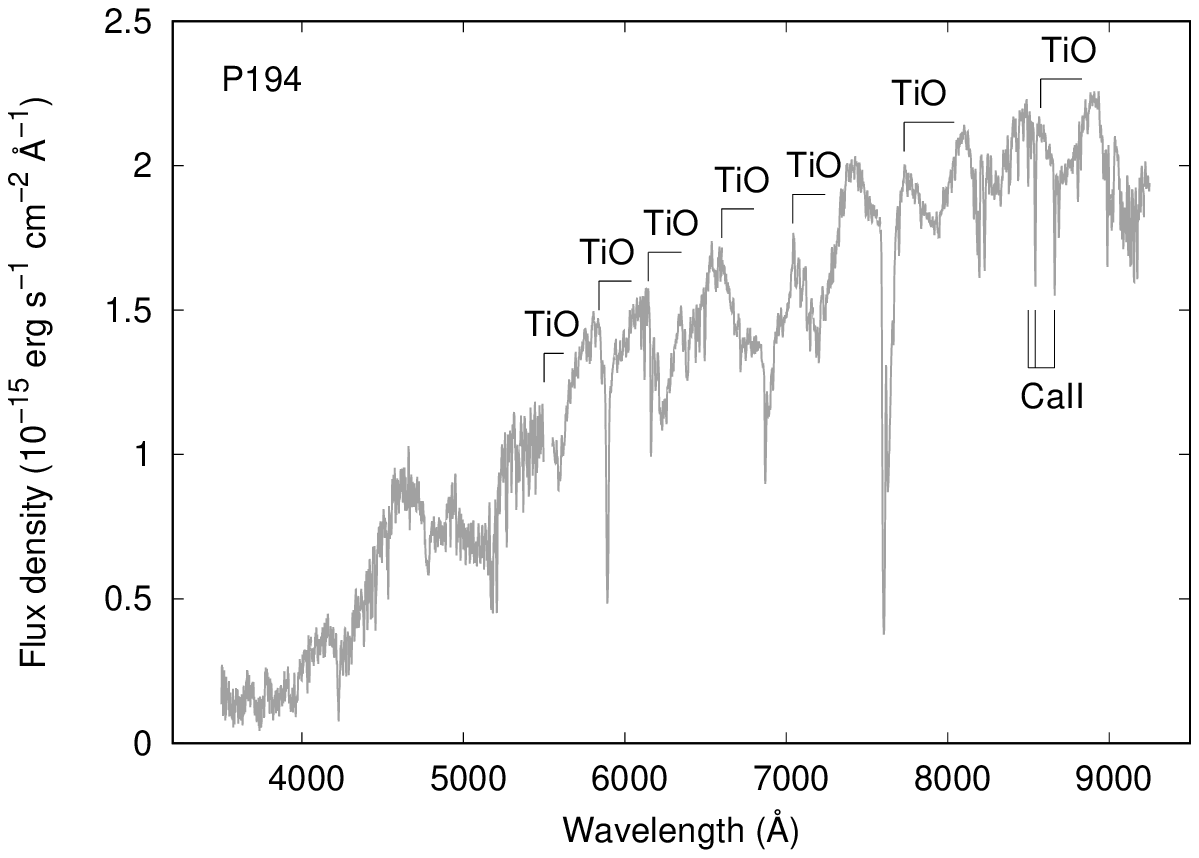}
\end{minipage}
\begin{minipage}[l]{0.47\textwidth}
\includegraphics[width=\textwidth]{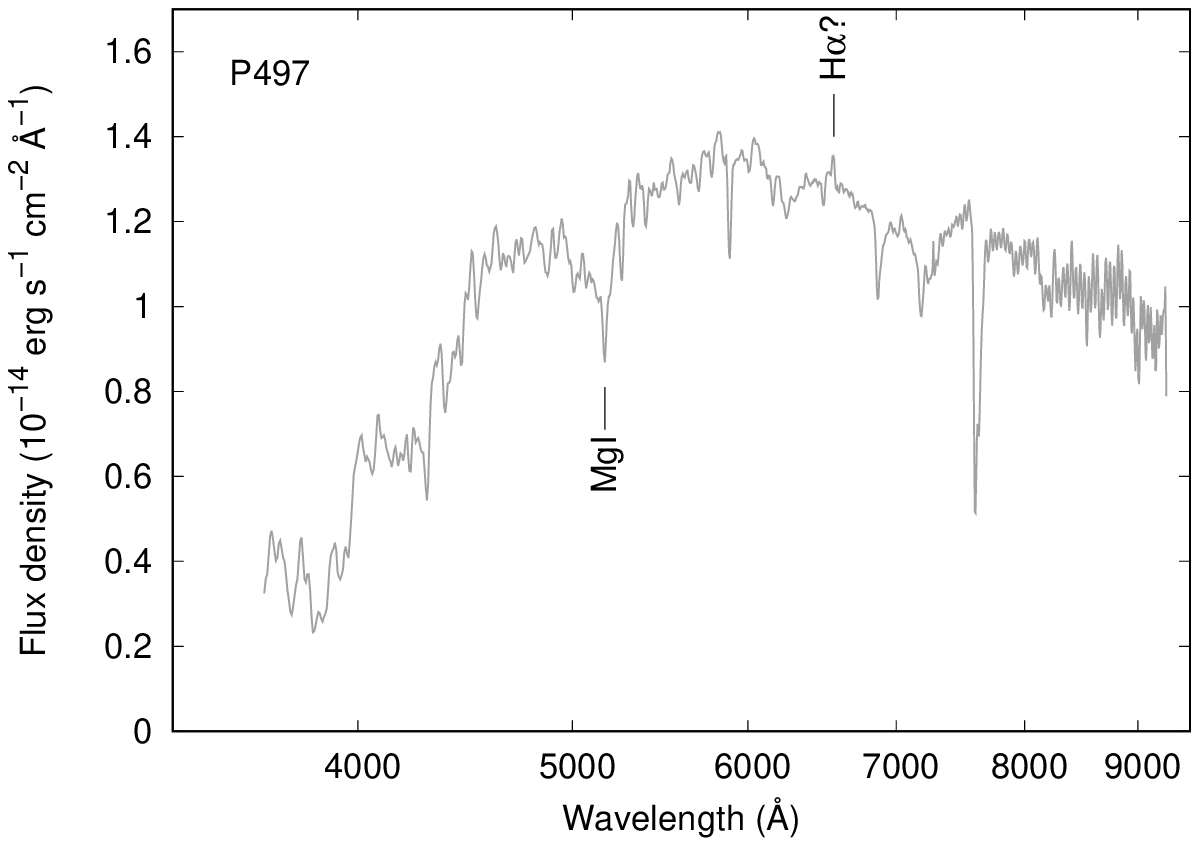}
\end{minipage}\\
\begin{minipage}[l]{0.47\textwidth}
\includegraphics[width=\textwidth]{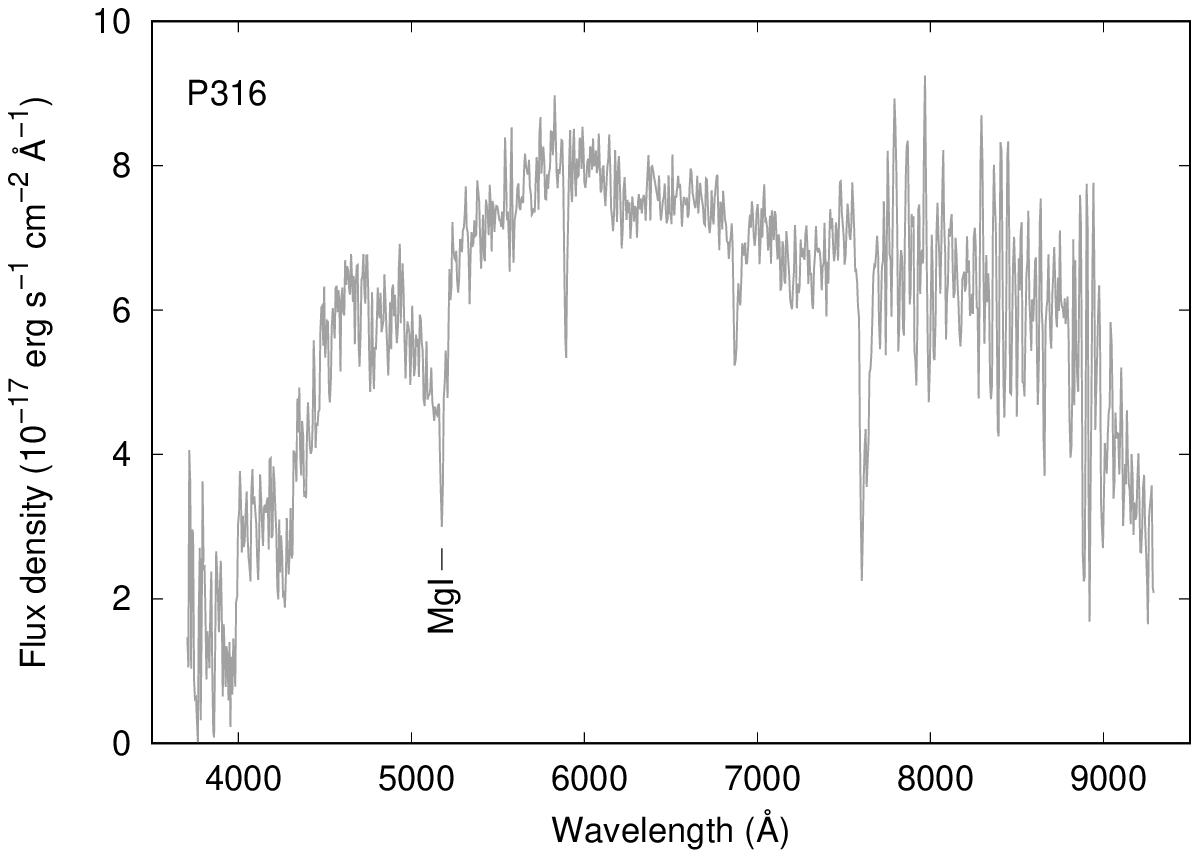}
\end{minipage}
\begin{minipage}[l]{0.47\textwidth}
\includegraphics[width=\textwidth]{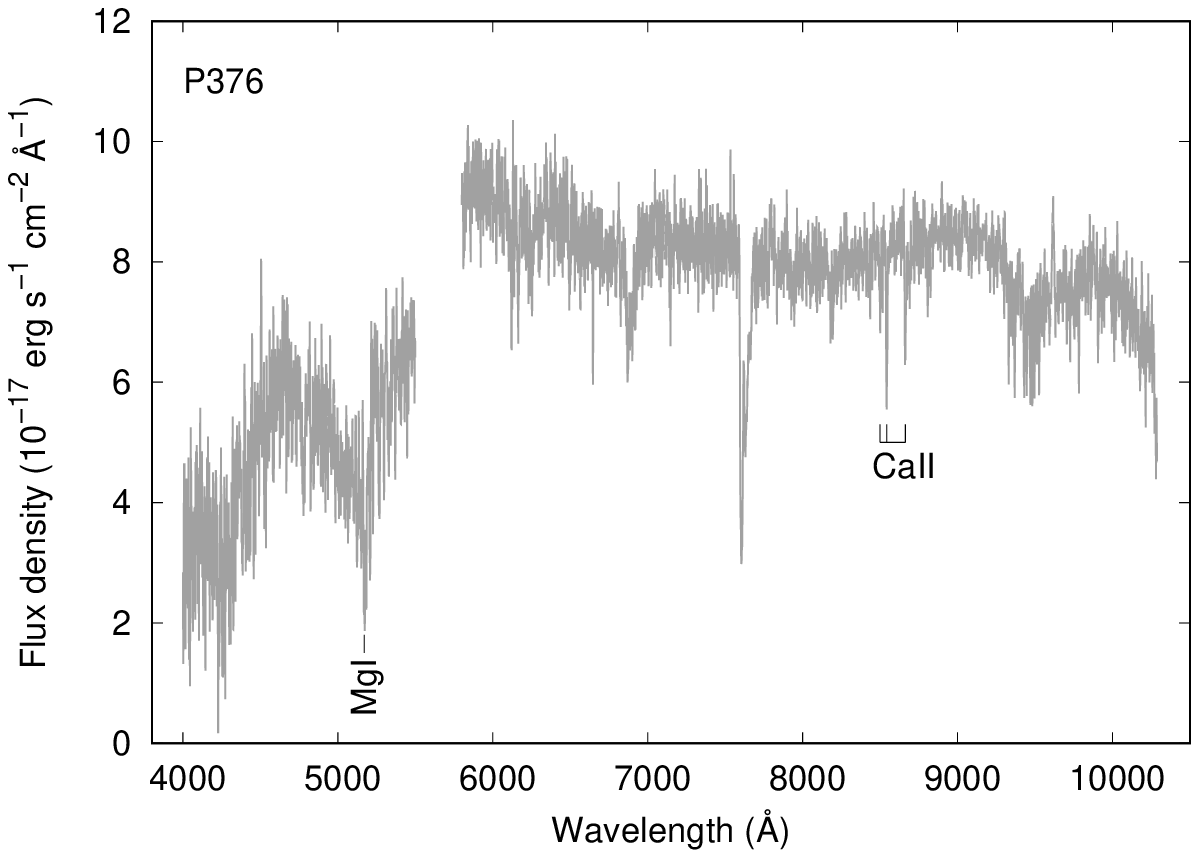}
\end{minipage}
\caption{The optical spectra for the sources that are dominated by a stellar component:
P127, P194, P497, P316, and P376.
The observational and data reduction information is provided in \cite{lansbury17}.
We have reassessed the line identifications, and they are labeled.}
\label{fig:opt_spec_from_farid_1}
\end{figure*}

\begin{figure*}
\centering
\begin{minipage}[l]{0.47\textwidth}
\includegraphics[width=\textwidth]{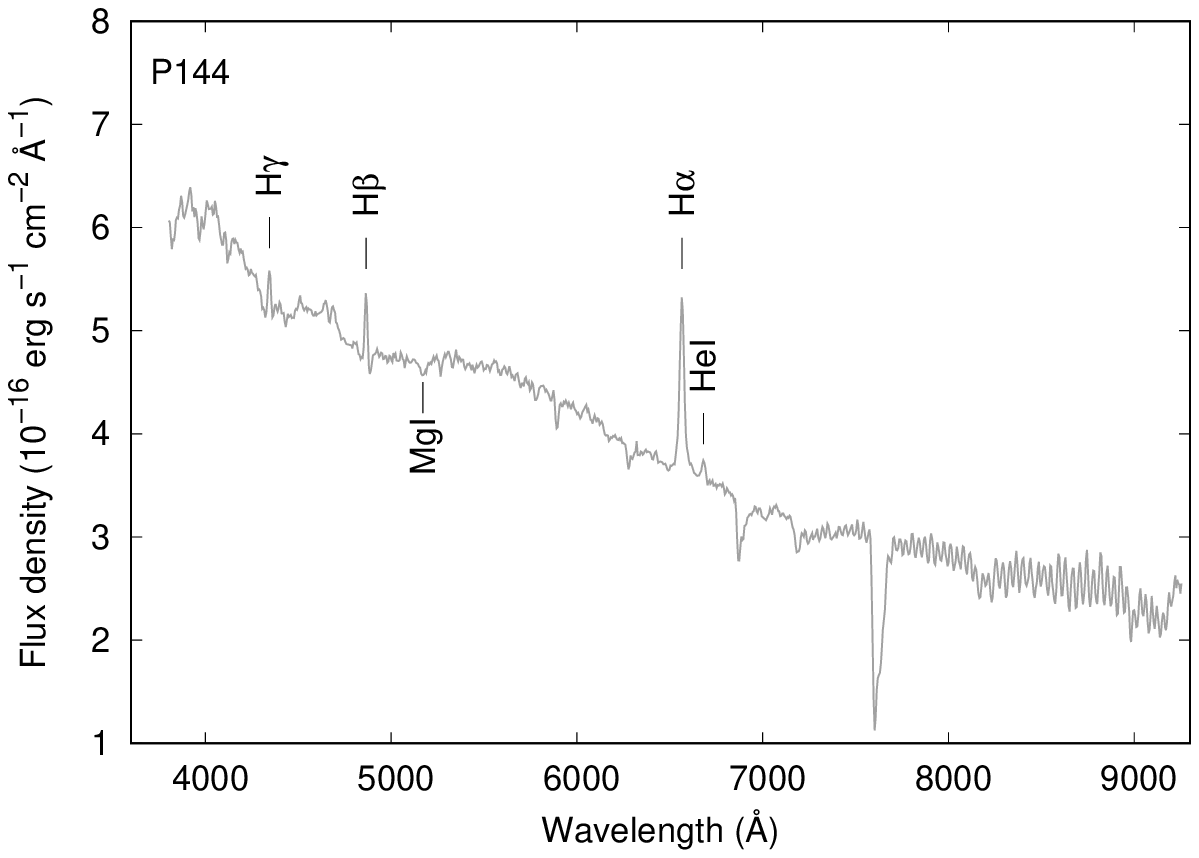}
\end{minipage}
\begin{minipage}[l]{0.47\textwidth}
\includegraphics[width=\textwidth]{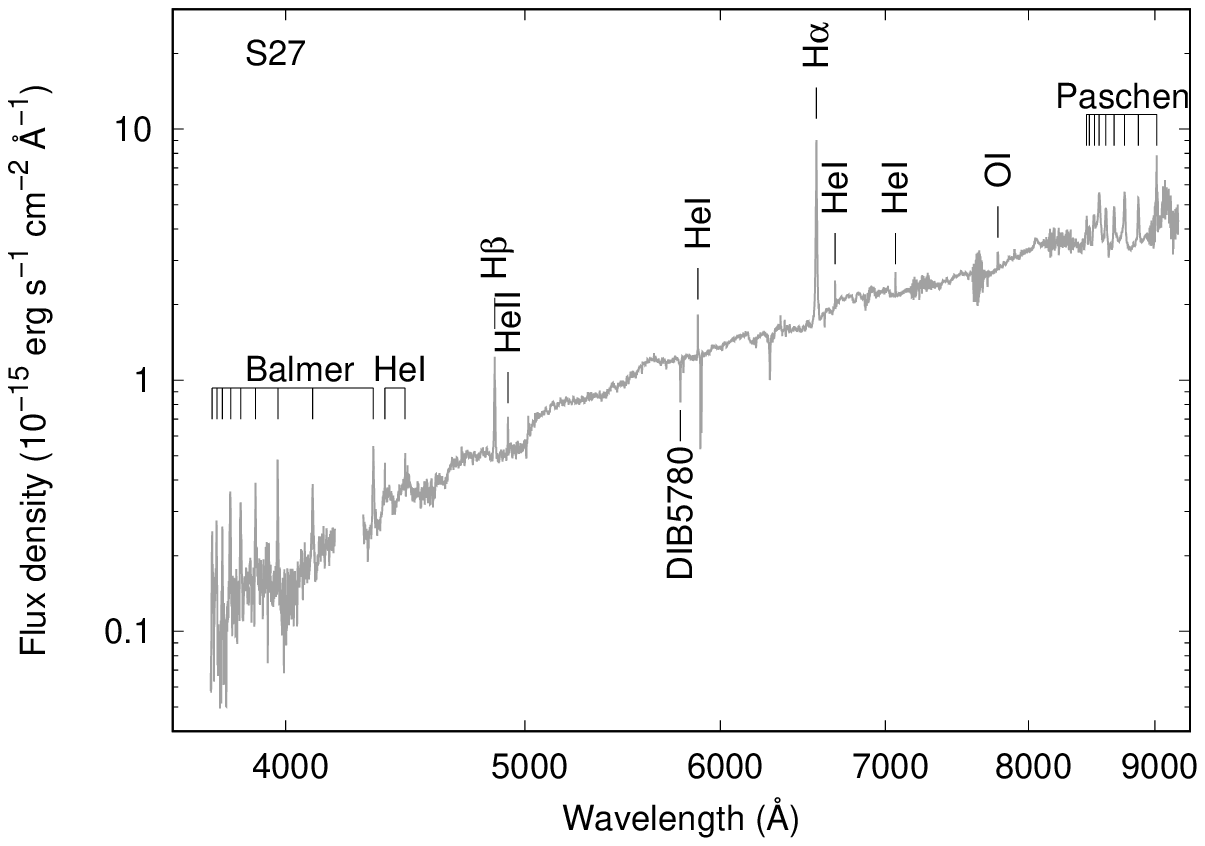}
\end{minipage}
\begin{minipage}[l]{0.47\textwidth}
\includegraphics[width=\textwidth]{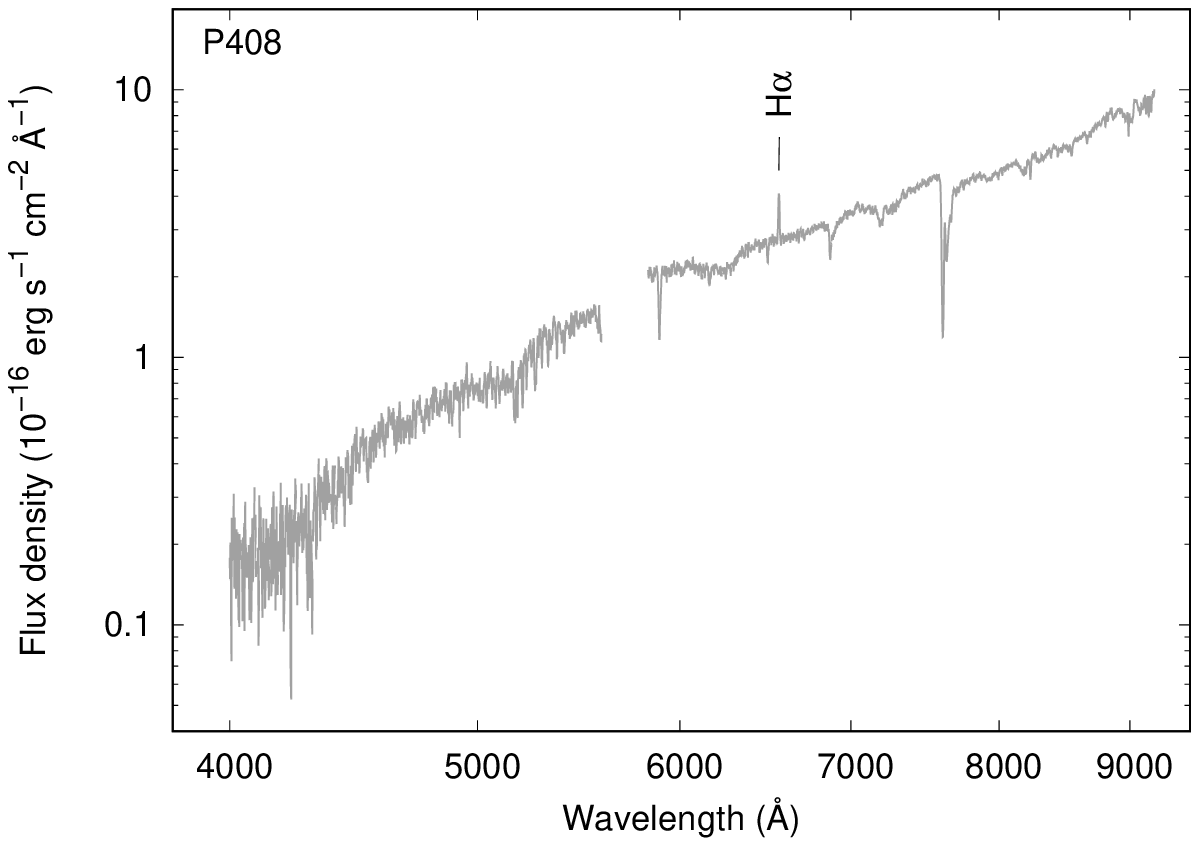}
\end{minipage}
\begin{minipage}[l]{0.47\textwidth}
\includegraphics[width=\textwidth]{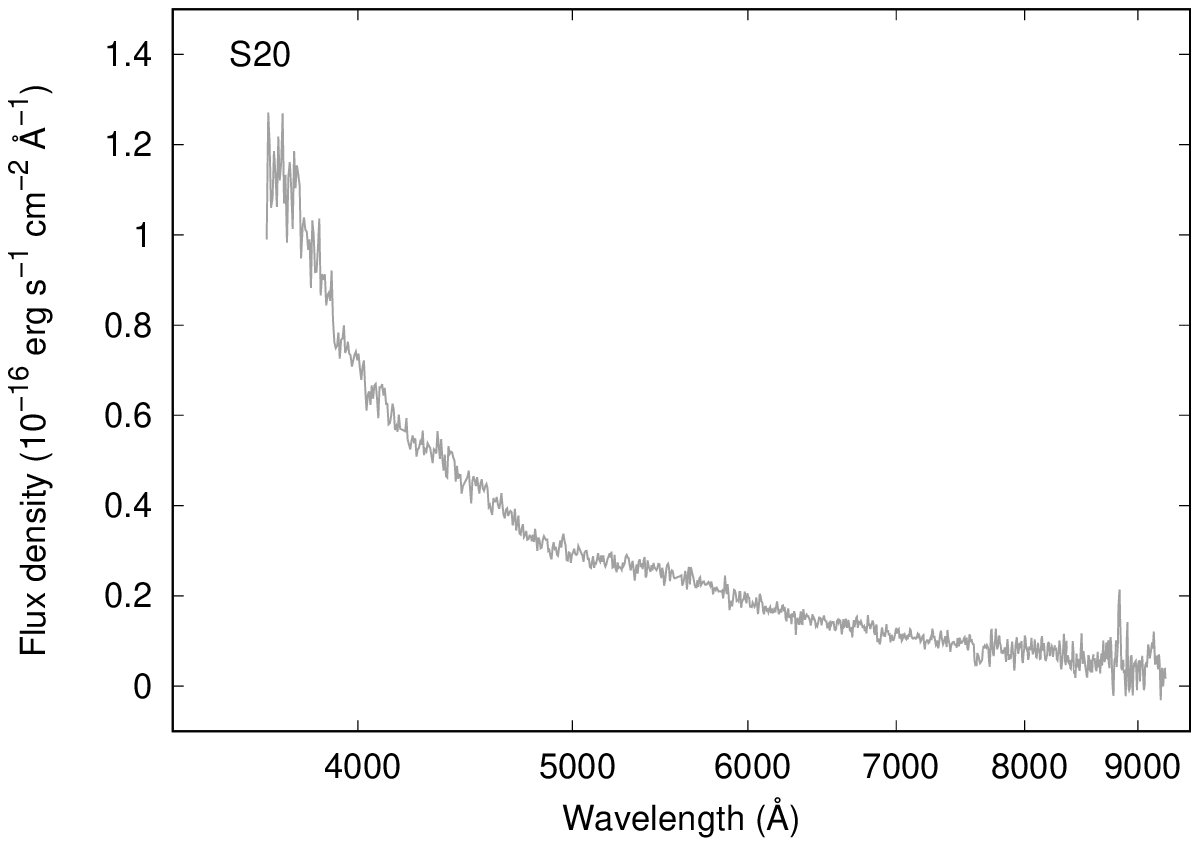}
\end{minipage}
\caption{The optical spectra for the sources that are dominated by a disk (accretion or
circumstellar): P144, S27, P408, and S20. The observational and data reduction information 
is provided in \cite{lansbury17}.  We have reassessed the line identifications, and they 
are labeled. }
\label{fig:opt_spec_from_farid_2}
\end{figure*}

Figures~\ref{fig:opt_spec_from_farid_1} and \ref{fig:opt_spec_from_farid_2} show 
the optical spectra from \cite{lansbury17} for the nine serendips that were not 
classified in the SIMBAD search described in Section 3.  As detailed in
\cite{lansbury17}, the observations occurred at four different telescopes between
2013 December 5 and 2016 February 13.  The spectra of P127, P144, P316, P497, and
S20 are from the New Technology Telescope (NTT), where the ESO Faint Object
Spectrograph and Camera (EFOSC2) was used.  P376 and P408 were observed from the
Keck telescope with the Low Resolution Imaging Spectrometer (LRIS).  The spectrum
of P194 came from the Double Spectrograph (DBSP) at Palomar, and S27's spectrum
was obtained with the Magellan Echellette (MagE) instrument.  Here, we consider how
the properties of the optical spectra as well as the X-ray properties constrain the
nature of the sources.  The spectra that appear to be dominated by stars are shown
in Figure~\ref{fig:opt_spec_from_farid_1}, and those that appear to be dominated by
a disk (accretion or circumstellar) are shown in Figure~\ref{fig:opt_spec_from_farid_2}.

The optical spectra of serendips P127 and P194 are dominated by absorption lines
from a stellar photosphere, and spectral types can be estimated.  For P127,
the Mg\,Ib absorption line at 5172\,\r{A}~is present but relatively weak, while
the Balmer absorption lines are strong.  These features suggest a mid-range spectral
type, and we estimate that it may be an F-type star.  Serendip P194 clearly shows
the TiO bands as well as a very red continuum, and both of these characteristics
are diagnostic of an M-type star.  While it is clear that the strongest contribution
to the optical spectra of these serendips is the star, it is unclear whether the
X-rays come from stellar coronal activity, either from an isolated star or an active
binary \citep[AB, e.g.,][]{dempsey93,fpt03} or if there is an accreting binary companion
(e.g., a white dwarf).  For P127, the only evidence that there might be an accreting
companion is the fact that it is detected in the 3--8\,keV band.  The evidence for an
accretor is a little stronger for P194 because the {\em NuSTAR} spectrum is relatively
hard, and the source is detected in the 8--24\,keV band (see Figure~\ref{fig:hf}).
It is also possible that P194 is a symbiotic binary with a giant star and a compact
object as it is now well-established that many symbiotics with white dwarf companions
produce X-ray emission \citep[e.g.,][]{luna13}.  However, in our case, we cannot
conclusively determine the luminosity class of the M-type star in P194.  We conclude
that serendips P127 and P194 are isolated stars, ABs, or CVs, and we note that
symbiotics with white dwarfs would fall in the class of CVs.

Serendip P497 appears to be intermediate in temperature between P127 and P194.
Although extinction is very low in its direction ($b$ = --72$^{\circ}$), it has a
redder continuum than P127 and also a stronger Mg\,Ib line.  Thus, serendip P497
appears to be dominated by a star with a K or G spectral type.  While serendip
P497 is only detected in the 3--8\,keV band and the X-ray spectrum is relatively
soft (see Figure~\ref{fig:hf}), there is some evidence in the optical spectrum
for an H$\alpha$ emission line.  The presence of an optical emission line could
indicate a contribution from an accretion disk, suggesting the possibility that
this source is a CV.  However, some ABs show H$\alpha$ in emission \citep{montes97},
so this is also a possibility.

Although the statistical quality of the optical spectra for serendips P316 and
P376 are lower, they also appear to be dominated by stellar photospheres.  The
optical spectrum of P316 looks extremely similar to P497 in terms of the redness
of the continuum and the strength of the Mg\,Ib line.  The Mg\,Ib line has
approximately the same strength in P376; thus, P316, P376, and P497 may all have
K or G spectral types.  Serendips P316 and P376 are only detected in the 3--8\,keV
band, but the constraints on the hardness of their X-ray spectra are weak
(see Figure~\ref{fig:hf}).  These two serendips also could be isolated stars,
ABs, or CVs.

Serendips P144, S27, and P408 all have Balmer emission lines in their optical
spectra, and Table~\ref{tab:opt_lines} lists the central wavelengths, the
equivalent widths, and the fluxes for the optical emission lines detected.
The lines provide evidence for the presence of an accretion disk (and thus a
compact companion) or circumstellar material (and possibly a compact companion).
For P144, the optical spectrum also has a Mg\,Ib line, suggesting that it
does not harbor a high-mass star.  Thus, we suggest that this source is
either a CV or an LMXB.  However, serendips S27 and P408 are both HMXB candidates.  
They are located in the Galactic plane with $b$ = --0.6$^{\circ}$ and $2.3^{\circ}$,
respectively, and both show strong extinction.  S27 has a very strong H$\alpha$ 
emission line, which is suggestive of a Be star and possibly a Be X-ray binary 
with a neutron star or a black hole.  For P408, the H$\alpha$ line is weaker, 
but the continuum is very similar to S27.  While S27 is detected in the 
8--24\,keV band, serendip P408 is not; thus, S27 is a somewhat stronger 
HMXB candidate.

Finally, S20 is unique among this group in having a very blue spectrum and no
narrow emission or absorption features.  Although weak lines might be uncovered
if a spectrum with higher signal-to-noise were obtained, the existing spectrum
suggests that S20's optical emission is dominated by thermal emission from an
accretion disk.  If the spectrum does lack hydrogen emission lines, then one 
possible explanation might be that the donor star is a hydrogen-poor white dwarf.  
In this scenario, the system might be an ultracompact X-ray binary (UCXB), and 
we note that the optical spectrum of S20 is very similar to the UCXB 4U~1246--58
\citep{intzand08}.  S20 also has very low extinction, which is not surprising 
given the fact that it is somewhat away from the Galactic plane at $b = 13^{\circ}$.
Its distance and luminosity are difficult to constrain, but the fact that its
8--24\,keV flux is high ($\sim$$5\times 10^{-11}$\,erg\,cm$^{-2}$\,s$^{-1}$) may
require the presence of a neutron star or black hole accretor.  Also, the dominance
of the accretion disk in the optical indicates that the donor star is not
luminous and that the source is not an HMXB.  The source may be an LMXB, but
we cannot rule out the possibility that it is a CV.  

\section{X-ray Spectroscopy for the Serendips with Optical Emission Lines}

The serendips that are most likely to contain compact objects are those exhibiting
accretion disk signatures in their optical spectra.  Considering our nine serendips 
in this context, there are three types.  The first type includes S20, which has an 
optical spectrum with no lines at all, and we argue above that we are likely seeing 
thermal emission from a hydrogen-poor accretion disk.  The second type includes sources 
with optical absorption lines but no emission lines.  Although these sources could 
have compact objects, they could also be isolated stars or ABs.  The third type is 
serendips with optical emission lines, and these lines are most likely to originate 
in an accretion disk around a compact object or in a circumstellar disk around a 
Be star.  Although isolated Be stars and possibly Be white dwarf systems exist, a
major class of HMXBs are Be X-ray binaries.

Here, we focus on the third type because it may be possible to use
the {\em NuSTAR} and archival {\em XMM-Newton} spectra to distinguish
between the CV, LMXB, and HMXB possibilities.  We consider the three
sources that have clear optical emission lines (P144, S27, and P408)
as well as P497, which may have weak H$\alpha$ in emission.  Most CVs 
that emit hard X-rays are likely to have X-ray spectra that are dominated 
by an optically thin plasma emitting thermal bremsstrahlung \citep{krivonos07_cv}.  
Although expectations for LMXBs depend on whether the compact object is
a neutron star or a black hole both have distinguishing X-ray features:
most quiescent neutron star LMXBs have a low temperature ($\sim$0.1\,keV)
blackbody component, and quiescent black hole LMXBs simply have a
power-law spectrum with a photon index of $\Gamma\sim 1.5$--2 \citep{plotkin13}.
Most HMXBs host highly magnetized neutron stars with hard X-ray spectra
$\Gamma\sim 1$.

We used data from the {\em NuSTAR} ObsIDs listed in Table~\ref{tab:obs} to 
produce 3--79\,keV energy spectra for Focal Plane Modules A and B (FPMA and 
FPMB).  Using {\ttfamily nuproducts}, we extracted source spectra from circular 
regions with $45^{\prime\prime}$ radii centered on the positions given in 
Table~\ref{tab:list}.  Given that the serendips are in the fields of relatively 
bright sources, the background for the serendip includes photons from the PSF 
wings of the bright sources as well as the normal instrumental background.  To 
estimate background, we extracted counts from an annulus centered on the bright 
target source.  We set the inner and outer radii of the annulus to match the 
serendip source region but removed any parts of the annulus within $100^{\prime\prime}$ 
of the serendip.  In addition to the source and background spectra, the
{\ttfamily nuproducts} routine produces the instrument response files.  For S27, 
there are multiple observations, and we combined spectra from different ObsIDs 
using {\ttfamily addspec}.  

To extend the coverage to lower energies, we searched the {\em XMM-Newton} 
archive for observations that include the serendips with optical emission lines 
(P144, S27, P408, and P497) in their fields of view.  For the four sources, 
there are, respectively, 1, 8, 1, and 2 observations.
Information about the {\em XMM} observations that we used is provided in
Table~\ref{tab:obs}.  We used all the available data for P144, P408, and P497,
and the longest of the eight observations for S27.  We analyzed the data from
the EPIC/pn instrument, which covers the 0.3--12\,keV bandpass and has the
highest effective area of the {\em XMM-Newton} instruments \citep{struder01}.
We used the Science Analysis Software (SAS) to extract source spectra from a
circular aperture with a radius of $20^{\prime\prime}$ for the first three sources
and a radius of $30^{\prime\prime}$ for P497.  The fact that P497 is brighter
is the reason that the larger radius is used.  We extracted background spectra
from a source-free rectangular region in another part of the field of view, and
then used {\ttfamily rmfgen} and {\ttfamily arfgen} to make the instrument response
files.

As indicated in Table~\ref{tab:obs}, the {\em XMM} and {\em NuSTAR} observations 
for P144, S27, P408, and P497 were separated by 7, 2, 4, and 0.3 years, respectively.  
Thus, as the {\em XMM} and {\em NuSTAR} bands overlap, this analysis also provides
information about the long-term X-ray variability of these sources.  While the
{\em XMM} and {\em NuSTAR} observations were not simultaneous, the {\em Swift}
satellite obtained soft X-ray coverage that was near-simultaneous with the {\em NuSTAR}
observations.  We have also produced spectra from the {\em Swift} X-ray Telescope
\citep{burrows05}, and the ObsIDs and exposure times used are listed in Table~\ref{tab:obs}.
Due to the relatively short observations, the fact that the sources are faint, and
the smaller effective area of {\em Swift}, the statistical quality of the data is
low.  Thus, we use them only as a check on source variability.

\begin{figure*}
\epsscale{1.1}
\plotone{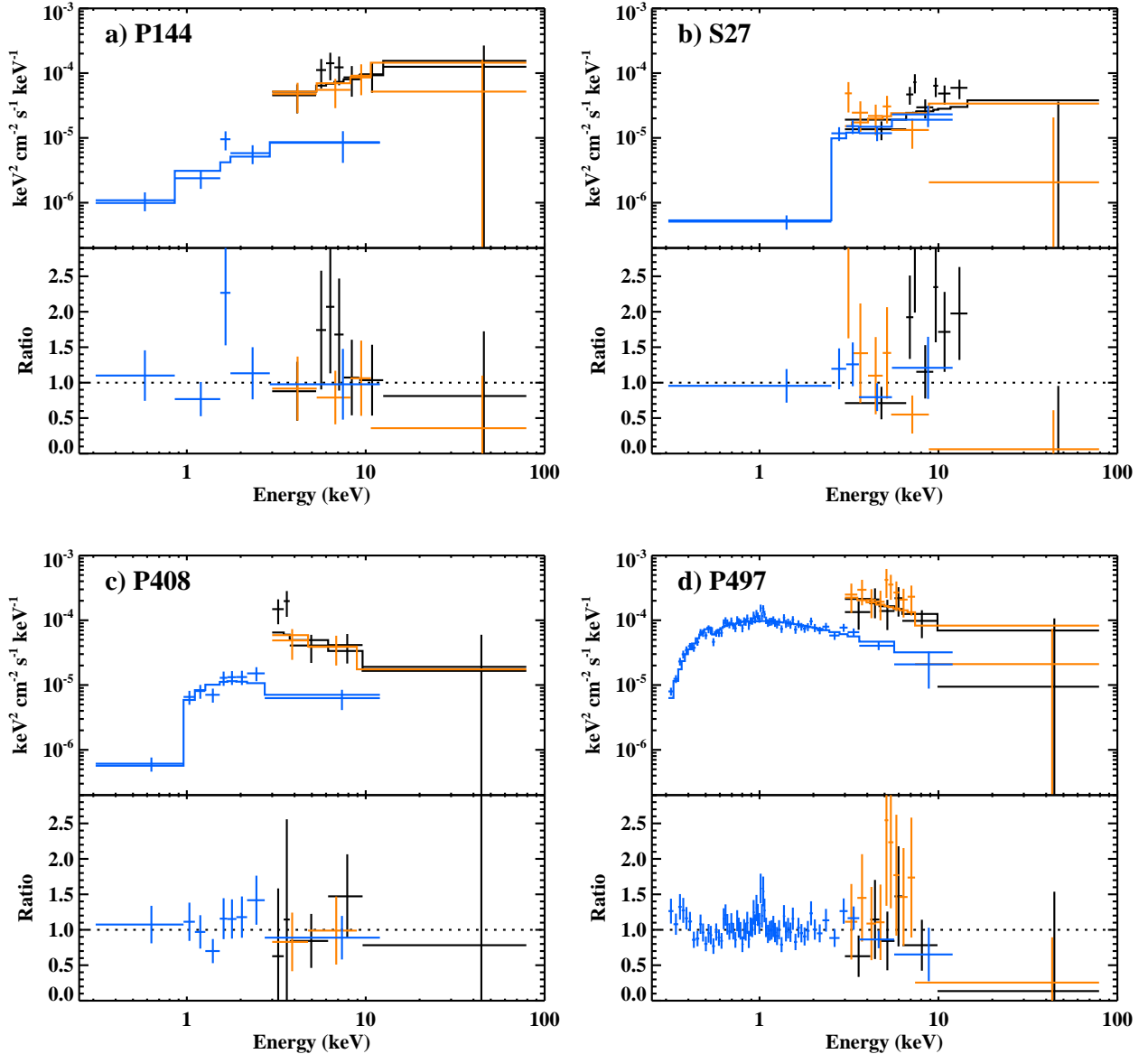}
\caption{X-ray spectra and data-to-model ratios for the four Galactic serendips 
with optical emission lines fitted with an absorbed power-law model.  The blue 
points are from observations by the pn instrument on {\em XMM}, the black points
are from {\em NuSTAR}/FPMA, and the orange points are from {\em NuSTAR}/FPMB.
\label{fig:x_spec}}
\end{figure*}

We used the XSPEC v12.9.0n software package to fit the {\em NuSTAR} and {\em XMM}
spectra, starting with a simple absorbed power-law model.  To model the absorption,
we used the {\ttfamily tbabs} model with \cite{wam00} abundances and \cite{vern96}
cross sections.  Although FPMA and FPMB typically have normalizations that are
different by a few percent, our spectra do not have high enough statistical quality
to be sensitive to differences at this level, and we set the FPMA/FPMB
cross-normalization parameter to be 1.0.  However, due to possible variability, we
allowed the overall pn normalization to be different than {\em NuSTAR}, and this
is given as $C_{XMM}/C_{NuSTAR}$ in Table~\ref{tab:parameters}.  The power-law model
provides a good description of the P144 and P408 spectra and somewhat worse fits
to the S27 and P497 spectra.  Despite the lower quality of the fits for the latter
two sources, the power-law parameters demonstrate that P144 and S27 have intrinsically
hard spectra with photon indices of $\Gamma = 1.4^{+0.5}_{-0.4}$ and $1.7^{+0.6}_{-0.5}$,
respectively (90\% confidence errors), compared to $2.9^{+0.6}_{-0.5}$ and $2.7\pm 0.1$
for P408 and P497, respectively.  The difference in hardness is clear from an
inspection of the energy spectra (Figure~\ref{fig:x_spec}).  The spectra also reveal
that the column density is highest for S27, intermediate for P408, and lowest for P144
and P497.  We also report values for fits with an absorbed bremsstrahlung model in
Table~\ref{tab:parameters}.  The differences in the $\chi^{2}$ values are not
significant, meaning that none of the spectra allow us to distinguish between the
two models.  The high temperatures for P144 and S27 simply confirm that these are
hard spectra, while the lower values of $\sim$2\,keV for P408 and P497 may indicate
that we are seeing thermal emission at the measured temperatures.

For P497, the spectrum shows positive residuals between 0.9 and 1.2\,keV, and adding
a Gaussian improves the fit statistic significantly from $\chi^{2}/\nu = 91/83$ to
$\chi^{2}/\nu = 70/80$.  The parameters of the Gaussian are constrained to
$E_{\rm line} = 1.01\pm 0.03$\,keV, $\sigma_{\rm line} < 0.09$\,keV, and
$N_{\rm line} = (2.5^{+1.5}_{-1.1})\times 10^{-5}$\,ph\,cm$^{-2}$\,s$^{-1}$.  The equivalent
width (EW) of the feature is $\sim$60\,eV.  The presence of emission lines in this
regime is consistent with the interpretation of the spectrum as being from an
optically thin thermal plasma with a temperature near 2\,keV as strong emission
lines from Fe\,XXII, Fe\,XXIII, Fe\,XXIV, and Ne\,X are all expected.  An iron line
near 6.4--6.7\,keV might also be anticipated, but the quality of the spectrum is not
sufficient to determine whether such a line is present.

Figure~\ref{fig:x_spec} shows strong variability for P144, P408, and P497, and this
is confirmed by the $C_{XMM}/C_{NuSTAR}$ values.  P144 and P408 were brighter by a
factor of $\sim$7 when {\em NuSTAR} observed them, and P497 was brighter by 4--5
times.  To confirm the interpretation that this is caused by long-term variability,
we added the near-simultaneous {\em Swift}/XRT data and refit the spectra with the
absorbed power-law model.  The values of $C_{Swift}/C_{NuSTAR}$ are given in
Table~\ref{tab:parameters}, and they are consistent with unity for P144, P408, and
P497.  Suprisingly, S27, which is the one source that did not show evidence for
long-term variability, has a value of $C_{Swift}/C_{NuSTAR} < 0.77$ (90\% confidence
upper limit).  As the {\em Swift} and {\em NuSTAR} observations were not strictly
simultaneous, this may indicate that S27 shows short-term variability.

Considering the information from the optical and X-ray spectra, it is likely
that P497 is a CV.  This classification is based on the evidence for a low-mass
companion star in the optical spectrum along with the evidence that the X-ray
emission is coming from a plasma with a temperature of $\sim$2\,keV.  Especially
with the weak optical emission lines, an AB nature is still a possibility for 
P497.  However, ABs usually show evidence for two temperature thermal emission, 
and they often have low coronal metallicity, making it less likely that an 
emission line at $\sim$1\,keV would be present \citep{fpt03}.  For S27,
the hardness of the X-ray spectrum is consistent with an HMXB nature, so this
source remains an HMXB candidate (further evidence in favor of an HMXB nature
is discussed in Section 7.2).  For P408, the X-ray continuum is very similar to
P497, suggesting a CV nature.  Although the P408 optical continuum matches best
with S27, this is probably due to the extinction being similar for the two
sources rather than indicating that they are intrinsically similar.  While we
cannot rule out an X-ray binary nature for P408, we consider it more likely to
be a CV.  P144 is a good candidate for being an LMXB.  If the source is in
quiescence, then the fact that we do not see a low temperature blackbody
component in the spectrum, which would be expected for neutron star LMXB,
may suggest that it is a black hole LMXB.  However, it is also possible that
P144 is a CV with a high bremsstrahlung temperature.

\section{Discussion}

Table~\ref{tab:classifications} provides a summary of the classifications or
possible classifications for all 16 of the serendips.  Although there is some
uncertainty about the classification for many of the sources, the identifications
include:  8 stars (although some of these may have binary companions); 4 CVs or
CV candidates (P82 and P98 are confirmed, P497 is likely, and P408 is a candidate);
2 LMXB candidates (P144 and S20); an HMXB (S43); and an HMXB candidate (S27).
In the following, we discuss these groups and the implications for the faint end
of the Galactic hard X-ray source population.  We also discuss the fact that,
at this stage, the survey is incomplete, and there are biases against some
source types.

\subsection{Stars, CVs, and LMXBs}

For the 8 stars, the classifications either come from the SIMBAD database
(for S1, P146, S37, and P340) or the optical spectra we show in this work
(for P127, P194, P316, and P376).  However, especially because they are X-ray
emitters, we cannot rule out the possibility that some of them have binary
companions, which could either be another star (i.e., they may be active
binaries, ABs) or a white dwarf (WD).  In three cases, parallax distances are available
in the {\em Gaia} Data Release 1 catalog \citep{gaia1,gaia2}:  S1, P146, and
P340 have distances of $33.8\pm 0.3$\,pc, $671\pm 117$\,pc, and $93\pm 4$\,pc,
respectively.  Thus, S1 and P340 are very nearby, and they may not require a
binary companion to produce the observed X-rays.  For S1, the 3--8\,keV flux 
measured by {\em NuSTAR} is $<$$3\times 10^{-14}$\,erg\,cm$^{-2}$\,s$^{-1}$, which
corresponds to a luminosity of $L < 4\times 10^{27}$\,erg\,s$^{-1}$.  For P340,
the 3--8\,keV flux is $6\times 10^{-14}$\,erg\,cm$^{-2}$\,s$^{-1}$, and the
luminosity is $L = 7\times 10^{28}$\,erg\,s$^{-1}$.  On the other hand, for
P146, the 3--8\,keV flux of $5\times 10^{-14}$\,erg\,cm$^{-2}$\,s$^{-1}$ implies
$L = 3\times 10^{30}$\,erg\,s$^{-1}$, which either requires an early spectral
type (i.e., a high mass star) or a binary companion.

In the strict sense of this study as a 8--24\,keV survey, we only detect one
star (P194), but it should be noted that P194 is in a part of the sky that
received a large amount of exposure time (216\,ks), and all of the other stars
have 8--24\,keV flux upper limits which are higher than the P194 flux.  Despite
this difference in exposure time, the fact remains that we know that P194 is a
hard X-ray emitter.  While some isolated M-type stars may produce soft X-ray
emission \citep{hunsch98}, the hard X-ray emission from P194 (see
Figure~\ref{fig:hf}) probably indicates the presence of a binary companion.
In Section 5, we discussed the possibility that P194 is a symbiotic.  Such
systems are known to produce hard X-ray emission \citep{kennea09}, and there
are also cases where the optical emission lines from such systems are very
weak \citep{mukai16}.

For the CVs, P82 was previously known to be the soft X-ray source RX~J0425.6--5714.
Being a polar-type, the spin of its white dwarf is synchronized to its orbit, and
the period is 1.43\,hr \citep{rk03}.  As polars are generally relatively soft X-ray
sources, it may be somewhat surprising that {\em NuSTAR} strongly detects the
source at $(7.5\pm 0.2)\times 10^{-13}$\,erg\,cm$^{-2}$\,s$^{-1}$ in the 8--24\,keV
band (Figure~\ref{fig:hf}); however, even higher-energy emission from polars is
not unprecedented \citep{barlow06}.  Given the high Galactic latitude of
$b$ = --42$^{\circ}$, the source may be relatively nearby, but no distance estimate
is available for RX~J0425.6--5714.  P98 is the CV V1193~Ori with an orbital period
of 3.96\,hr.  Prior to {\em NuSTAR}, the only X-ray information published on
V1193~Ori was a very weak {\em ROSAT} detection of its soft X-ray flux
\citep{verbunt97}.  Its distance is constrained to be $>$470\,pc \citep{ringwald94};
thus, the 3--8\,keV flux of $(1.21\pm 0.10)\times 10^{-13}$\,erg\,cm$^{-2}$\,s$^{-1}$
corresponds to a luminosity of $>$$3.2\times 10^{30}$\,erg\,s$^{-1}$.  For P497,
the evidence that the X-ray spectrum is due to a $\sim$2\,keV thermal plasma
makes it likely that this source is a CV.  The optical spectrum is dominated
by a K or G type star, allowing for us to make a rough distance estimate.  We
assume that the spectral type is K0\,V, which indicates an absolute magnitude
of $M_{V} = 5.9$ \citep{cox00}.  The column density from the X-ray spectrum is
$N_{\rm H} = 3.1\times 10^{20}$\,cm$^{-2}$ (Table~\ref{tab:parameters}), and this
corresponds to $A_{V} = 0.14$ \citep{go09}.  Table~\ref{tab:list_opt} lists an
$R$-band magnitude of 12.49, and $V$--$R = 0.74$ for a K0\,V star, leading to
an estimate of $V = 13.2$, and we calculate a distance of $d\sim 270$\,pc.
Given the Galactic latitude of $b$ = --72$^{\circ}$, a much larger distance than
this would be surprising \citep{revnivtsev08}.  Thus, the unabsorbed 2--10\,keV
flux of $4.8\times 10^{-13}$\,erg\,cm$^{-2}$\,s$^{-1}$ (Table~\ref{tab:parameters})
corresponds to a luminosity of $\sim$$4\times 10^{30}$\,erg\,s$^{-1}$, which is
reasonable for a CV, but is also possible for an AB \citep{sazonov06}.

While P144 and S20 may also be CVs, they are the only sources in this study
for which an LMXB nature is equally (or perhaps marginally more) plausible.
Although their optical spectra are different in that P144 has emission lines
and S20 does not, we argue that both are dominated by emission from an accretion
disk (possibly due to S20 being an UCXB transferring hydrogen-poor material).  
However, as we do not know the sizes of the accretion disks, the absolute
magnitudes of P144 and S20 are unknown.  For P144, the upper limit on the measured 
column density is $<$$4.6\times 10^{21}$\,cm$^{-2}$, which is little help in 
constraining the distance to the source since the column density through the Galaxy 
along the line of sight to P144 ($l = 246.45^{\circ}$, $b$ = --4.66$^{\circ}$) is
$4.4\times 10^{21}$\,cm$^{-2}$ \citep{kalberla05}.  If P144 was a black hole LMXB
in outburst, the accretion disk would account for all of the optical light;
however, the fact that we see the Mg\,Ib line suggests that there is still a
small contribution from the companion, and the putative black hole LMXB may be
in or near quiescence.  Black hole LMXBs in quiescence typically have X-ray
luminosities between $10^{30}$ and $10^{33}$\,erg\,s$^{-1}$ \citep{garcia01}.
For this range of luminosities, the unabsorbed 2--10\,keV flux that we measure
(Table~\ref{tab:parameters}) corresponds to a distance range of 0.2--7.5\,kpc.
While not at all constraining, this does show that P144 may plausibly be a
quiescent black hole LMXB.  As the vast majority of black hole LMXBs that we
know of were discovered in outburst, the possibility that hard X-ray surveys
may be able to find such systems in quiescence is highly significant.  If S20
is a black hole or neutron star LMXB, it is certainly not in quiescence,
and we will report on details of dedicated {\em XMM} and {\em NuSTAR}
observations in a future paper.

\subsection{HMXBs}

A main result of this work is the discovery of the HMXB candidate S27,
NuSTAR~J105008--5958.8.  Although to this point, we have focused on the
HMXB evidence from the optical emission lines and the hard X-ray spectrum,
another important feature is the diffuse interstellar band (DIB)
5780\,\AA~absorption line (Figure~\ref{fig:opt_spec_from_farid_2}).  The
EW of the line is $0.97\pm 0.10$\,\AA, which corresponds to
$E(B-V) = 1.50\pm 0.15$ \citep{1994Jenniskens} and, using $A_{V} = 3.1 E(B-V)$,
an optical extinction of $A_{V} = 4.7\pm 0.5$.  For this extinction and at
the location of the source ($l = 288.30^{\circ}$, $b$ = --0.60$^{\circ}$),
we use the \cite{marshall06} extinction maps to estimate a distance to
S27 of 6--8\,kpc.  Although the source is $0.7^{\circ}$ from the Carina
nebula, which is at a distance of 2.3\,kpc, the larger distance for
S27 indicates that they are not associated.

The distance estimate also allows us to determine the absolute optical
magnitude and the X-ray luminosity.  Convolving the flux values shown
in Figure~\ref{fig:opt_spec_from_farid_2} with $R$-band and $V$-band
filter profiles gives $R = 15.1$, which is near the USNO-B1.0 value
(see Table~\ref{tab:list_opt}), and $V = 16.5$.  Combining this with
the $A_{V}$ value and a distance of $7\pm 1$\,kpc gives
$M_{V}$ = --$2.4\pm 0.6$, which is the absolute magnitude for the star
and the circumstellar material combined.  For Be X-ray binaries, there
is a relationship between the H$\alpha$ EW and the excess emission from
the circumstellar material \citep{rtn12}, and, for the S27 value of
--$28.2\pm 4.4$\,\AA, the correction is 0.3 magnitudes at $V$-band.
Thus, for the star alone, we estimate $M_{V}$ = --$2.1\pm 0.6$, which
is consistent with main sequence star classifications between B2 and B3
\citep{cox00} and our previous suggestion that the companion is a Be star.
Concerning the X-rays, at a 2--10\,keV unabsorbed flux of
$(5.8^{+1.9}_{-1.5})\times 10^{-14}$\,erg\,cm$^{-2}$\,s$^{-1}$
and the 6--8\,kpc distance range, the source luminosity is between
$1.8\times 10^{32}$\,erg\,s$^{-1}$ and $5.9\times 10^{32}$\,erg\,s$^{-1}$.
These luminosity values are higher than those that are seen for isolated
B-type stars \citep[e.g.,][]{rauw15}, which, along with the fact that S27
is detected in the 8--24\,keV band, strongly favors the presence of a
compact object in the system.  However, we still consider S27 to be an
HMXB candidate (rather than a certain HMXB) because we cannot necessarily
rule out the possibility that it is a colliding wind binary.  Also, we
consider below whether S27 may be a $\gamma$\,Cas analog \citep{shrader15,motch15}.

\begin{figure}
\hspace{-0.5cm}
\epsscale{1.2}
\plotone{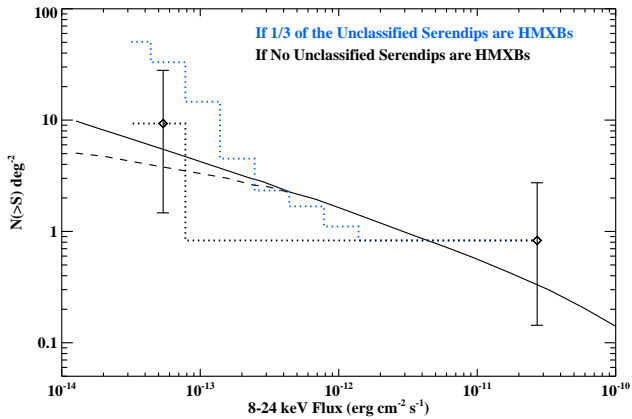}  
\caption{The surface density vs. 8--24\,keV flux (i.e., the log$N$-log$S$)
for HMXBs.  The dashed line is a prediction for wind-fed persistent HMXBs
from \cite{lutovinov13}, and the solid line is an extrapolation from the
curves at higher fluxes, which have been previously measured down to
$\sim$$1\times 10^{-11}$\,erg\,cm$^{-2}$\,s$^{-1}$.  The points and their 
68\% confidence Poisson errors correspond to the HMXB S43 and the HMXB 
candidate S27.  There are 24 serendips within $5^{\circ}$ of the Galactic 
plane detected in the 8--24\,keV band, which have not been classified.  The 
dotted black line represents a possible surface density if none of them are 
HMXBs, and the dotted blue line shows the surface density if eight of them 
are HMXBs.\label{fig:predict}}
\end{figure}

As constraining the faint end of the HMXB population is a goal of the
{\em NuSTAR} Galactic surveys \citep{harrison13}, we discuss our HMXB
results in the larger context of the luminosity function and surface
density (log$N$-log$S$) for HMXBs in the Galaxy.  \cite{lutovinov13}
show that, with the {\em INTEGRAL} survey being complete for persistent
HMXBs down to $\sim$$10^{-11}$\,erg\,cm$^{-2}$\,s$^{-1}$ (17--60\,keV), the
surface density is relatively well-constrained down to this level
(although it is not uniform across the Galaxy).  However, there is
significant uncertainty below this level.  Figure~\ref{fig:predict}
shows predictions for the log$N$-log$S$ \citep[from][]{lutovinov13}
for what we might see by extending the HMXB search to lower flux levels.
The dashed curve corresponds to a case where the luminosity curve
flattens below $10^{34}$\,erg\,s$^{-1}$ as predicted for wind-fed
accretion in HMXBs due to the fact that the minimum in the stellar
mass distribution ($\sim$8--10\Msun~for HMXBs) leads to a minimum in
the mass transfer rate from the wind \citep{castor75} and, thus, the
luminosity \citep{lutovinov13}.  At a flux of
$10^{-14}$\,erg\,cm$^{-2}$\,s$^{-1}$, this causes a drop in
the number of HMXBs per square degree by a factor of $\sim$2 over a
simple extrapolation of the slope at higher luminosities.  In fact,
the \cite{lutovinov13} model leaves out other physics, such as the
possible impact of the propeller mechanism \citep{is75}, which could
cause the surface density to drop even lower than the dashed curve.
Although the curves from \cite{lutovinov13} are for the 17--60\,keV
band, we converted them to 8--24\,keV using the spectral parameters
reported in \cite{coburn02} for ten HMXBs.  The mean of the 8--24\,keV
fluxes is 1.23 times larger than the mean of the 17--60\,keV fluxes,
and we shifted the curves by that amount.

In our study, S43 is a definite HMXB \citep[e.g.,][]{krivonos15}
with an 8--24\,keV flux of $2.71\times 10^{-11}$\,erg\,cm$^{-2}$\,s$^{-1}$,
and S27 is a strong HMXB candidate with an 8--24\,keV flux of
$5.4\times 10^{-14}$\,erg\,cm$^{-2}$\,s$^{-1}$ (Table~\ref{tab:parameters}).
Although these are only two sources, we can still use this information
to make a log$N$-log$S$ plot and compare it to the predictions.  We use
the same approach that we used for our earlier {\em Chandra} study of the
Norma spiral arm region.  Equation 15 of \cite{fornasini14} depends on
the sky coverage as a function of flux.  Given that HMXBs are strongly
clustered in the Galactic plane, we use the sky coverage within $5^{\circ}$
of the plane (see Figure~\ref{fig:coverage}).  We assume that the
probability functions are delta functions at the fluxes of S27 and S43.
The Poisson errors are much larger than the uncertainty introduced by
these approximations.  Figure~\ref{fig:predict} compares the log$N$-log$S$
for the two sources with 68\% confidence Poisson errors to the curves from
\cite{lutovinov13}, and the measurements are consistent with both curves.

One factor that must be considered in the interpretation of the log$N$-log$S$
is the incompleteness of the source classifications.  In total, there are
30 serendips within $5^{\circ}$ of the Galactic plane detected in the 8--24\,keV
band.  However, only six of these have been classified, including S27 and S43.
Thus, we have produced completeness-corrected log$N$-log$S$ curves using the
fluxes of the 24 unclassified sources, and making two different assumptions
about how many of them are HMXBs.  One possibility is that none of them are
HMXBs, leaving just S27 and S43 (see Figure~\ref{fig:predict}).  Motivated
by the prevalence of HMXBs in the classified group (two of six), we also
consider the possibility that 1/3 of the unclassified serendips (eight of
24) are HMXBs, and that is also shown in Figure~\ref{fig:predict}.  To produce
the blue dotted curve, we used the following, which is similar to Equation 15
from \cite{fornasini14},
\begin{equation}
N(>f_{x}) = \int_{f_{x}}^{\infty}\left[\sum_{i=1}^{26}\frac{P_{i}(f_{x})}{A(f_{x})}\right]~df_{x},
\end{equation}
where $P_{i}$ are the probabilities that each source is an HMXB.  Thus, 
$P_{i} = 1$ for the 8--24\,keV fluxes, $f_{x}$, of S27 and S43, and 
$P_{i} = 1/3$ for the 8--24\,keV fluxes of the other 24 sources.  Also, 
$A(f_{x})$ is the sky coverage for $\left| b\right| < 5^{\circ}$ as shown in 
Figure~\ref{fig:coverage}.  The resulting curve indicates that the surface 
density of faint HMXBs below $\sim$$10^{-13}$\,erg\,cm$^{-2}$\,s$^{-1}$
could be several times higher than the \cite{lutovinov13} predictions.
However, if none or only a couple of the unclassified sources
are HMXBs, then the surface density may be consistent with the predictions.
Determining the nature of a large fraction of the 24 unclassified 
serendips would provide a definitive comparison.

In addition to classifying more of the serendips in the Galactic plane,
it is also important to learn more about S27 (NuSTAR~J105008--5958.8).
The HMXBs considered by \cite{lutovinov13} in their study of the surface
density and luminosity function were persistent systems.  While many of
the \cite{lutovinov13} HMXBs are persistent Be systems, a large fraction
of the known Be systems are transient.  In quiescence, the transient
Be systems can have either hard power-law or soft blackbody spectra
\citep{tsygankov17}.  Thus, while the fact that S27 has a power-law
spectrum does not distinguish between it being transient or persistent,
it may be an indication that accretion onto a compact object is occurring.

S27 has some properties in common with the $\gamma$\,Cas binary 
system.  $\gamma$\,Cas as well as about ten $\gamma$\,Cas analogs
produce hard X-ray emission, but its origin is still debated. $\gamma$\,Cas
is composed of a Be star in 203.6\,day orbit with a white dwarf or possibly
a neutron star.  The hard X-ray emission may come from accretion onto the
compact object, but it may also be produced by magnetic interactions between
the Be star and its circumstellar disk \citep{shrader15,motch15}.  In the
neutron star scenario, the low X-ray luminosity may be explained if the
rotation rate of the neutron star is high enough to centrifugally inhibit
accretion.  \cite{postnov17} recently considered such a scenario for
$\gamma$\,Cas where the X-rays come from a hot shell of material accumulated
outside the neutron star's magnetosphere.

\subsection{Survey Incompleteness}

In considering the results related to Galactic hard X-ray populations, we must
keep in mind that this study has only included sources with identifications
via optical spectroscopy.  For the full coverage area, of the 497 detected
{\em NuSTAR} sources (in the 3--24\,keV band for the primary catalog only),
identifications were obtained for 276 sources, making the completeness
fraction 56\%.  The main reasons why sources are not identified include:
1. the possibility that the {\em NuSTAR} source is spurious; 2. optical
source confusion in crowded regions
\citep[indicated for many sources in Table~6 of][]{lansbury17}; 3. intrinsic
faintness in the optical; 4. faintness of sources in the optical due to
interstellar absorption; 5. lack of a soft X-ray counterpart, which could
either be due to variability or the lack of deep enough X-ray coverage.
While there is a 1--8\,ks {\em Swift}/XRT observation with nearly every
{\em NuSTAR} observation, the short XRT observations are not always deep
enough to detect the faint sources.

The completeness levels strongly depend on Galactic latitude from 63\% (261
identified out of 415) for sources that are more than $10^{\circ}$ away from the
plane to 32\% (8 identified out of 25) for sources that are 5--10$^{\circ}$ away,
to 12\% (7 identified out of 57) for sources within $5^{\circ}$ of the plane.
The low fraction of identified sources close to the plane is consistent with the
fact that there is more crowding and more extinction there.  As stars and CVs
are relatively nearby and are spread across latitudes, the incompleteness
fractions suggest that we might find a factor of $\sim$2 more of these with a
complete survey.  However, the actual number is probably smaller since isolated
stars that are bright enough to be detected by {\em NuSTAR} are very bright in
the optical and are not likely to be missed (unless there is a lack of soft
X-ray coverage), and CVs are also relatively bright in the optical.

At low Galactic latitudes, especially within $5^{\circ}$ of the plane, the
small completeness fraction (12\%) raises the question of what source types
remain unidentified in the Galactic plane.  Figure~\ref{fig:coverage} shows that
the serendip coverage in this region is very similar to the Norma coverage, and,
for Norma, we estimated that about eight of the detected sources are AGN
\citep{fornasini17}.  Another source type to consider is magnetars, which are
highly magnetized isolated neutron stars.  Magnetars are so faint in the optical
that there is no chance that any of these would be included in our study, but they
could be detected by {\em NuSTAR}.  In fact, we do know that one of the sources
in the \cite{lansbury17} catalog, NuSTAR~J183452--0845.6 (P420), is the magnetar
Swift~J1834.9--0846.  Finally, as discussed in Section 7.2, a major question 
for our study is how many HMXBs we might be missing.

\section{Conclusions and Future Work}

The results presented here give a first systematic look at the Galactic sources
that are being found in the {\em NuSTAR} serendipitous survey.  As we have
included 16 sources from all Galactic latitudes, it is not surprising that a 
relatively large fraction (11 out of 16) of the sources are stars or CVs.  
In addition, as this is a study of Galactic hard X-ray populations,
we consider all detected sources other than the target of the observation, and
this leads to the inclusion of some previously known sources such as the HMXB
2RXP~J130159.6--635806 and the CVs V1193~Ori and RX~J0425.6--5714.  However,
the survey has also uncovered new sources.  NuSTAR~J073959--3147.8 (P144) is
an LMXB candidate with an X-ray spectrum that is well-described by a relatively
hard power-law.  If the system is a quiescent LMXB, the lack of a thermal
blackbody component in the spectrum favors a black hole accretor over a neutron
star.  It is an important development if {\em NuSTAR} can help us pick out
quiescent BH systems since there should be a large number of these.
NuSTAR~J092418--3142.2 (S20) is the brightest serendip, but the source has
never been studied previously.  We conclude that it is either an LMXB
(possibly a UCXB) or a CV, and we will report on a dedicated {\em XMM-Newton}
and {\em NuSTAR} observation of the source in an upcoming paper.  

The discovery of the HMXB candidate NuSTAR~J105008--5958.8 (S27) is especially
interesting because of the possibility that there is a large population of
faint HMXBs in the Galaxy.  In addition to further studies of this source,
potentially to determine its orbital period, the classifications of the
{\em NuSTAR} serendips are especially incomplete in the Galactic plane, and
determining the nature of the 24 serendips within $5^{\circ}$ of the plane
would allow us to determine if the surface density of HMXBs is consistent
with the predictions of \cite{lutovinov13}.  The work on searching for
HMXBs in the Norma region is also on-going.  The sky coverage for Norma is
very similar to the Serendipitous survey coverage in the Galactic plane
(see Figure~\ref{fig:coverage}), and there are three HMXB candidates in
the Norma region \citep{rahoui14,fornasini17}.  The nature of these HMXB
candidates still requires confirmation, and we have an approved near-IR
spectroscopy program to search for orbital motion.  With continued effort
on classifying sources from both the Serendipitous and Norma surveys, it
will be possible to combine the results and stongly constrain the surface
density of HMXBs in the Galaxy.

\acknowledgments

This work made use of data from the {\it NuSTAR} mission, a project led by the
California Institute of Technology, managed by the Jet Propulsion Laboratory,
and funded by the National Aeronautics and Space Administration. We thank the
{\it NuSTAR} Operations, Software and  Calibration teams for support with the
execution and analysis of these observations.  This research has made use of
the {\it NuSTAR}  Data Analysis Software (NuSTARDAS) jointly developed by the
ASI Science Data Center (ASDC, Italy) and the California Institute of Technology
(USA).  Data from {\em Swift}, {\em XMM}, and {\em Chandra} were also used, and 
the work on serendipitous {\em NuSTAR} sources is partially funded by {\em Chandra}
grants GO5-16154X and GO6-17135X.  RK acknowledges support from Russian Science
Foundation (grant 14-12-01315).  We acknowledge useful discussions with A.~Shaw,
and we thank the referee for helpful comments.
This work has made use of data from the European Space Agency (ESA) mission 
{\em Gaia}, processed by the {\em Gaia} Data Processing and Analysis Consortium 
(DPAC).  Funding for the DPAC has been provided by national institutions, in 
particular the institutions participating in the {\em Gaia} Multilateral Agreement.  
This research had made use of the SIMBAD database and the VizieR catalog access 
tool, CDS, Strasbourg, France. 


\begin{thebibliography}{}

\bibitem[\protect\astroncite{{Abbott} et~al.}{2016}]{abbott16}
{Abbott}, B.~P., {Abbott}, R., {Abbott}, T.~D., et~al.\  2016, Physical Review
  Letters, 116, 061102

\bibitem[\protect\astroncite{{Aird} et~al.}{2015}]{aird15}
{Aird}, J., {Alexander}, D.~M., {Ballantyne}, D.~R., et~al.\  2015, ApJ, 815,
  66

\bibitem[\protect\astroncite{{Ajello} et~al.}{2012}]{ajello12}
{Ajello}, M., {Alexander}, D.~M., {Greiner}, J., et~al.\  2012, ApJ, 749, 21

\bibitem[\protect\astroncite{{Alexander} et~al.}{2013}]{alexander13}
{Alexander}, D.~M., {Stern}, D., {Del Moro}, A., et~al.\  2013, ApJ, 773, 125

\bibitem[\protect\astroncite{{Barlow} et~al.}{2006}]{barlow06}
{Barlow}, E.~J., {Knigge}, C., {Bird}, A.~J., et~al.\  2006, MNRAS, 372, 224

\bibitem[\protect\astroncite{{Bird} et~al.}{2016}]{bird16}
{Bird}, A.~J., {Bazzano}, A., {Malizia}, A., et~al.\  2016, ApJS, 223, 15

\bibitem[\protect\astroncite{{Bond} et~al.}{1987}]{bond87}
{Bond}, H.~E., {Grauer}, A.~D., {Burstein}, D., \& {Marzke}, R.~O.  1987, PASP,
  99, 1097

\bibitem[\protect\astroncite{{Brorby} et~al.}{2016}]{brorby16}
{Brorby}, M., {Kaaret}, P., {Prestwich}, A., \& {Mirabel}, I.~F.  2016, MNRAS,
  457, 4081

\bibitem[\protect\astroncite{{Brown} et~al.}{2016}]{gaia1}
{Brown}, A.~G.~A., {Vallenari}, A., {Prusti}, T., et~al.\  2016, A\&A, 595, A2

\bibitem[\protect\astroncite{{Burrows} et~al.}{2005}]{burrows05}
{Burrows}, D.~N., {Hill}, J.~E., {Nousek}, J.~A., et~al.\  2005, Space Science
  Reviews, 120, 165

\bibitem[\protect\astroncite{{Castor} et~al.}{1975}]{castor75}
{Castor}, J.~I., {Abbott}, D.~C., \& {Klein}, R.~I.  1975, ApJ, 195, 157

\bibitem[\protect\astroncite{{Civano} et~al.}{2015}]{civano15}
{Civano}, F., {Hickox}, R.~C., {Puccetti}, S., et~al.\  2015, ApJ, 808, 185

\bibitem[\protect\astroncite{{Coburn} et~al.}{2002}]{coburn02}
{Coburn}, W., {Heindl}, W.~A., {Rothschild}, R.~E., et~al.\  2002, ApJ, 580,
  394

\bibitem[\protect\astroncite{{Cox}}{2000}]{cox00}
{Cox}, A.~N.,  2000,
\newblock {Allen's astrophysical quantities},
\newblock  4th ed.~ Publisher: New York: AIP Press; Springer; Edited by Arthur
  N.~Cox.

\bibitem[\protect\astroncite{{Dempsey} et~al.}{1993}]{dempsey93}
{Dempsey}, R.~C., {Linsky}, J.~L., {Schmitt}, J.~H.~M.~M., \& {Fleming}, T.~A.
  1993, ApJ, 413, 333

\bibitem[\protect\astroncite{{Evans} et~al.}{2010}]{evans10}
{Evans}, I.~N., {Primini}, F.~A., {Glotfelty}, K.~J., et~al.\  2010, ApJS, 189,
  37

\bibitem[\protect\astroncite{{Evans} et~al.}{2014}]{evans14}
{Evans}, P.~A., {Osborne}, J.~P., {Beardmore}, A.~P., et~al.\  2014, ApJS, 210,
  8

\bibitem[\protect\astroncite{{Fornasini} et~al.}{2014}]{fornasini14}
{Fornasini}, F.~M., {Tomsick}, J.~A., {Bodaghee}, A., et~al.\  2014, ApJ, 796,
  105

\bibitem[\protect\astroncite{{Fornasini} et~al.}{2017}]{fornasini17}
{Fornasini}, F.~M., {Tomsick}, J.~A., {Hong}, J., et~al.\  2017, ApJS, 229, 33

\bibitem[\protect\astroncite{{Franciosini} et~al.}{2003}]{fpt03}
{Franciosini}, E., {Pallavicini}, R., \& {Tagliaferri}, G.  2003, A\&A, 399,
  279

\bibitem[\protect\astroncite{{Garcia} et~al.}{2001}]{garcia01}
{Garcia}, M.~R., {McClintock}, J.~E., {Narayan}, R., et~al.\  2001, ApJ, 553,
  L47

\bibitem[\protect\astroncite{{G{\"u}ver} \& {{\"O}zel}}{2009}]{go09}
{G{\"u}ver}, T., \& {{\"O}zel}, F.  2009, MNRAS, 400, 2050

\bibitem[\protect\astroncite{{Halpern} et~al.}{1998}]{halpern98}
{Halpern}, J.~P., {Leighly}, K.~M., {Marshall}, H.~L., {Eracleous}, M., \&
  {Storchi-Bergmann}, T.  1998, PASP, 110, 1394

\bibitem[\protect\astroncite{{Harrison} et~al.}{2013}]{harrison13}
{Harrison}, F.~A., {Craig}, W.~W., {Christensen}, F.~E., et~al.\  2013, ApJ,
  770, 103

\bibitem[\protect\astroncite{{Hong} et~al.}{2016}]{hong16}
{Hong}, J., {Mori}, K., {Hailey}, C.~J., et~al.\  2016, ApJ, 825, 132

\bibitem[\protect\astroncite{{Hunsch} et~al.}{1998}]{hunsch98}
{Hunsch}, M., {Schmitt}, J.~H.~M.~M., {Schroder}, K.-P., \& {Zickgraf}, F.-J.
  1998, A\&A, 330, 225

\bibitem[\protect\astroncite{{Illarionov} \& {Sunyaev}}{1975}]{is75}
{Illarionov}, A.~F., \& {Sunyaev}, R.~A.  1975, A\&A, 39, 185

\bibitem[\protect\astroncite{{in't Zand} et~al.}{2008}]{intzand08}
{in't Zand}, J.~J.~M., {Bassa}, C.~G., {Jonker}, P.~G., et~al.\  2008, A\&A,
  485, 183

\bibitem[\protect\astroncite{{Jenniskens} \& {Desert}}{1994}]{1994Jenniskens}
{Jenniskens}, P., \& {Desert}, F.-X.  1994, A\&AS, 106, 39

\bibitem[\protect\astroncite{{Kalberla} et~al.}{2005}]{kalberla05}
{Kalberla}, P.~M.~W., {Burton}, W.~B., {Hartmann}, D., et~al.\  2005, A\&A,
  440, 775

\bibitem[\protect\astroncite{{Kennea} et~al.}{2009}]{kennea09}
{Kennea}, J.~A., {Mukai}, K., {Sokoloski}, J.~L., et~al.\  2009, ApJ, 701, 1992

\bibitem[\protect\astroncite{{Krivonos} et~al.}{2007}]{krivonos07_cv}
{Krivonos}, R., {Revnivtsev}, M., {Churazov}, E., et~al.\  2007, A\&A, 463, 957

\bibitem[\protect\astroncite{{Krivonos} et~al.}{2012}]{krivonos12}
{Krivonos}, R., {Tsygankov}, S., {Lutovinov}, A., et~al.\  2012, A\&A, 545, A27

\bibitem[\protect\astroncite{{Krivonos} et~al.}{2015}]{krivonos15}
{Krivonos}, R.~A., {Tsygankov}, S.~S., {Lutovinov}, A.~A., et~al.\  2015, ApJ,
  809, 140

\bibitem[\protect\astroncite{{Lansbury} et~al.}{2017}]{lansbury17}
{Lansbury}, G.~B., {Stern}, D., {Aird}, J., et~al.\  2017, ApJ, 836, 99

\bibitem[\protect\astroncite{{Luna} et~al.}{2013}]{luna13}
{Luna}, G.~J.~M., {Sokoloski}, J.~L., {Mukai}, K., \& {Nelson}, T.  2013, A\&A,
  559, A6

\bibitem[\protect\astroncite{{Lutovinov} et~al.}{2013}]{lutovinov13}
{Lutovinov}, A.~A., {Revnivtsev}, M.~G., {Tsygankov}, S.~S., \& {Krivonos},
  R.~A.  2013, MNRAS, 431, 327

\bibitem[\protect\astroncite{{Marshall} et~al.}{2006}]{marshall06}
{Marshall}, D.~J., {Robin}, A.~C., {Reyl{\'e}}, C., {Schultheis}, M., \&
  {Picaud}, S.  2006, A\&A, 453, 635

\bibitem[\protect\astroncite{{Mineo} et~al.}{2012}]{mineo12}
{Mineo}, S., {Gilfanov}, M., \& {Sunyaev}, R.  2012, MNRAS, 419, 2095

\bibitem[\protect\astroncite{{Montes} et~al.}{1997}]{montes97}
{Montes}, D., {Fernandez-Figueroa}, M.~J., {de Castro}, E., \& {Sanz-Forcada},
  J.  1997, A\&AS, 125

\bibitem[\protect\astroncite{{Motch} et~al.}{2015}]{motch15}
{Motch}, C., {Lopes de Oliveira}, R., \& {Smith}, M.~A.  2015, ApJ, 806, 177

\bibitem[\protect\astroncite{{Mukai} et~al.}{2016}]{mukai16}
{Mukai}, K., {Luna}, G.~J.~M., {Cusumano}, G., et~al.\  2016, MNRAS, 461, L1

\bibitem[\protect\astroncite{{Mullaney} et~al.}{2015}]{mullaney15}
{Mullaney}, J.~R., {Del-Moro}, A., {Aird}, J., et~al.\  2015, ApJ, 808, 184

\bibitem[\protect\astroncite{{Plotkin} et~al.}{2013}]{plotkin13}
{Plotkin}, R.~M., {Gallo}, E., \& {Jonker}, P.~G.  2013, ApJ, 773, 59

\bibitem[\protect\astroncite{{Postnov} et~al.}{2017}]{postnov17}
{Postnov}, K., {Oskinova}, L., \& {Torrej{\'o}n}, J.~M.  2017, MNRAS, 465, L119

\bibitem[\protect\astroncite{{Pretorius} \& {Knigge}}{2012}]{pk12}
{Pretorius}, M.~L., \& {Knigge}, C.  2012, MNRAS, 419, 1442

\bibitem[\protect\astroncite{{Prusti} et~al.}{2016}]{gaia2}
{Prusti}, T., {de Bruijne}, J.~H.~J., {Brown}, A.~G.~A., et~al.\  2016, A\&A,
  595, A1

\bibitem[\protect\astroncite{{Rahoui} et~al.}{2014}]{rahoui14}
{Rahoui}, F., {Tomsick}, J.~A., {Fornasini}, F.~M., {Bodaghee}, A., \& {Bauer},
  F.~E.  2014, A\&A, 568, A54

\bibitem[\protect\astroncite{{Rauw} et~al.}{2015}]{rauw15}
{Rauw}, G., {Naz{\'e}}, Y., {Wright}, N.~J., et~al.\  2015, ApJS, 221, 1

\bibitem[\protect\astroncite{{Revnivtsev} et~al.}{2008}]{revnivtsev08}
{Revnivtsev}, M., {Sazonov}, S., {Krivonos}, R., {Ritter}, H., \& {Sunyaev}, R.
   2008, A\&A, 489, 1121

\bibitem[\protect\astroncite{{Ringwald} et~al.}{1994}]{ringwald94}
{Ringwald}, F.~A., {Thorstensen}, J.~R., \& {Hamwey}, R.~M.  1994, MNRAS, 271,
  323

\bibitem[\protect\astroncite{{Riquelme} et~al.}{2012}]{rtn12}
{Riquelme}, M.~S., {Torrej{\'o}n}, J.~M., \& {Negueruela}, I.  2012, A\&A, 539,
  A114
  
\bibitem[\protect\astroncite{{Ritter} \& {Kolb}}{2003}]{rk03}
{Ritter}, H., \& {Kolb}, U.  2003, A\&A, 404, 301

\bibitem[\protect\astroncite{{Rosen} et~al.}{2016}]{rosen16}
{Rosen}, S.~R., {Webb}, N.~A., {Watson}, M.~G., et~al.\  2016, A\&A, 590, A1

\bibitem[\protect\astroncite{{Sazonov} et~al.}{2006}]{sazonov06}
{Sazonov}, S., {Revnivtsev}, M., {Gilfanov}, M., {Churazov}, E., \& {Sunyaev},
  R.  2006, A\&A, 450, 117

\bibitem[\protect\astroncite{{Shrader} et~al.}{2015}]{shrader15}
{Shrader}, C.~R., {Hamaguchi}, K., {Sturner}, S.~J., et~al.\  2015, ApJ, 799,
  84

\bibitem[\protect\astroncite{{Str{\"u}der} et~al.}{2001}]{struder01}
{Str{\"u}der}, L., {Briel}, U., {Dennerl}, K., et~al.\  2001, A\&A, 365, L18

\bibitem[\protect\astroncite{{Sugizaki} et~al.}{2001}]{sugizaki01}
{Sugizaki}, M., {Mitsuda}, K., {Kaneda}, H., et~al.\  2001, ApJS, 134, 77

\bibitem[\protect\astroncite{{Tsygankov} et~al.}{2017}]{tsygankov17}
{Tsygankov}, S.~S., {Wijnands}, R., {Lutovinov}, A.~A., {Degenaar}, N., \&
  {Poutanen}, J.  2017, arXiv:1703.04634, submitted to MNRAS

\bibitem[\protect\astroncite{{Verbunt} et~al.}{1997}]{verbunt97}
{Verbunt}, F., {Bunk}, W.~H., {Ritter}, H., \& {Pfeffermann}, E.  1997, A\&A,
  327, 602

\bibitem[\protect\astroncite{{Verner} et~al.}{1996}]{vern96}
{Verner}, D.~A., {Ferland}, G.~J., {Korista}, K.~T., \& {Yakovlev}, D.~G.
  1996, ApJ, 465, 487

\bibitem[\protect\astroncite{{Voss} \& {Ajello}}{2010}]{va10}
{Voss}, R., \& {Ajello}, M.  2010, \apj, 721, 1843

\bibitem[\protect\astroncite{{Walter} et~al.}{2015}]{walter15}
{Walter}, R., {Lutovinov}, A.~A., {Bozzo}, E., \& {Tsygankov}, S.~S.  2015,
  A\&A~Rev., 23, 2

\bibitem[\protect\astroncite{{Wenger} et~al.}{2000}]{wenger00}
{Wenger}, M., {Ochsenbein}, F., {Egret}, D., et~al.\  2000, A\&AS, 143, 9

\bibitem[\protect\astroncite{{Wilms} et~al.}{2000}]{wam00}
{Wilms}, J., {Allen}, A., \& {McCray}, R.  2000, ApJ, 542, 914

\end{thebibliography}


\clearpage

\begin{table}
\caption{{\em NuSTAR} Galactic Serendips\label{tab:list}}
\begin{minipage}{\linewidth}
\begin{center}
\begin{tabular}{ccccccc} \hline \hline
Serendip & NuSTAR          & RA (J2000)\footnote{Except for S20, these are the {\em NuSTAR} positions from \cite{lansbury17}.  The 90\% confidence uncertainties are 14--22$^{\prime\prime}$, depending on the source detection significance.  For S20, a {\em Swift} position is given, and its uncertainty is 3.5$^{\prime\prime}$. } &   Decl (J2000)$^{a}$ & $l$        &   $b$   &  {\em NuSTAR}\\
ID       & Name            & (deg)      &   (deg)        & (deg)      &   (deg) &  Exposure (ks)\\ \hline\hline
S1       & J001639+8139.8  &   4.1662   &   81.6639      & 121.6049   &   18.8821 & 62\\
P82      & J042538--5714.5 &  66.4114   & --57.2417      & 267.0583   & --42.0332 & 243\\
P98      & J051626--0012.2 &  79.1107   &  --0.2047      & 201.7790   & --21.1032 & 148\\
P127     & J070014+1416.8  & 105.0607   &   14.2814      & 201.1168   &    8.3727 & 123\\
P144     & J073959--3147.8 & 114.9977   & --31.7969      & 246.4540   &  --4.6581 & 25\\
P146     & J075611--4133.9 & 119.0475   & --41.5666      & 256.5893   &  --6.7133 & 32\\
S20      & J092418--3142.2 & 141.07572  & --31.70479     & 259.5671   &   13.2248 & --\\
P194     & J095838+6909.2  & 149.6588   &   69.1538      & 141.7846   &   41.0720 & 216\\
S27      & J105008--5958.8  & 162.5362   & --59.9814      & 288.3008   &  --0.6010 & 348\\
S37      & J123559--3951.9  & 188.9960   & --39.8661      & 299.7133   &   22.9090 & 44\\
S43      & J130157--6358.1  & 195.4894   & --63.9699      & 304.0859   &  --1.1217 & 19\\
P316     & J133628--3414.1  & 204.1200   & --34.2350      & 313.4391   &   27.7160 & 132\\
P340     & J143636+5843.0  & 219.1532   &   58.7182      & 100.1750   &   53.5145 & 28\\
P376     & J165351+3938.5  & 253.4627   &   39.6424      &  63.4489   &   38.8521 & 18\\
P408     & J182604--0707.9  & 276.5178  &   --7.1331      &  23.6887   &    2.3455 & 31\\
P497     & J233426--2343.9  & 353.6111  &  --23.7331      &  39.7136   & --72.3073 & 15\\ \hline
\end{tabular}
\end{center}
\end{minipage}
\end{table}

\begin{table}
\caption{Soft X-ray Counterparts\label{tab:list_soft}}
\begin{minipage}{\linewidth}
\begin{center}
\begin{tabular}{ccccc} \hline \hline
Serendip & X-ray            & RA (J2000) &   Decl (J2000) & Separation\footnote{The angular distance between the positions reported in Table~\ref{tab:list}, which are {\em NuSTAR} positions except for S20, and the soft X-ray positions.}\\
ID       & Source           & (deg)      &   (deg)        & (arcsec)\\ \hline\hline
S1       & 3XMM J001652.0+813948  &   4.21681  &   81.66335     & 26.5\\
P82      & 3XMM J042538.6--571435 &  66.41093  & --57.24333     &  5.8\\
P98      & {\em Swift}\footnote{Here, we give the X-ray position from the \cite{lansbury17} catalog, which was determined from an analysis of {\em Swift}/XRT archival data.  However, we note that 1SXPS~J051626.6--001215 is a cataloged {\em Swift} source \citep{evans14} that is only $3^{\prime\prime}$ away from the \cite{lansbury17} position.} &  79.11136  &  --0.20499     &  2.8\\
P127     & 3XMM J070014.3+141644  & 105.05995  &   14.27906     &  8.8\\
P144     & 3XMM J074000.5--314759 & 115.00208  & --31.79997     & 17.5\\
P146     & CXO J075611.9--413358  & 119.04957  & --41.56628     &  5.8\\
S20      & CXO J092418.2--314217  & 141.07582  & --31.70497     &  0.7\\
P194     & 3XMM J095839.4+690910  & 149.66425  &   69.15279     &  7.9\\
S27      & 3XMM J105008.1--595902 & 162.53416  & --59.98389     &  9.7\\
S37      & CXO J123600.5--395215  & 189.00211  & --39.87102     & 24.5\\
S43      & 3XMM J130158.7--635808 & 195.49500  & --63.96917     &  9.2\\
P316     & 3XMM J133628.7--341356 & 204.11948  & --34.23235     &  9.8\\
P340     & 3XMM J143637.4+584303  & 219.15605  &   58.71761     &  5.8\\
P376     & 3XMM J165350.5+393821  & 253.46062  &   39.63944     & 12.1\\
P408     & CXO J182604.6--070806  & 276.51929  &  --7.13514     &  9.3\\
P497     & {\em Swift}\footnote{Here, we give the X-ray position from the \cite{lansbury17} catalog, which was determined from an analysis of {\em Swift}/XRT archival data.  However, we note that 1SXPS~J233426.6--234411 is a cataloged {\em Swift} source \citep{evans14} that is only $0.8^{\prime\prime}$ away from the \cite{lansbury17} position.} & 353.61102  & --23.73661     & 12.8\\ \hline
\end{tabular}
\end{center}
\end{minipage}
\end{table}

\begin{table}
\caption{Optical Counterparts\label{tab:list_opt}}
\begin{minipage}{\linewidth}
\begin{center}
\begin{tabular}{ccccccc} \hline \hline
Serendip & Optical     &   RA (J2000) &   Decl (J2000) & Separation\footnote{The angular distance between the soft X-ray and optical positions.} & $R$-band\\
ID       & Source      &   (deg)      &   (deg)        & (arcsec)   & magnitude\\ \hline\hline
S1       & USNO-B1.0 1716-0000986     &   4.21885    &   81.66364     & 1.48       &  8.16\\
P82      & USNO-B1.0 0327-0051610     &  66.41077    & --57.24344     & 0.50       & 18.21\\ 
P98      & SDSS J051626.67-001214.3   &  79.11113    &  --0.20400     & 3.67       & 13.98\\
P127     & USNO-B1.0 1042-0123735     & 105.06069    &   14.27916     & 2.61       & 17.41\\
P144     & USNO-B1.0 0582-0158974     & 115.00190    & --31.79980     & 0.84       & 17.11\\
P146     & USNO-B1.0 0484-0117243     & 119.04954    & --41.56631     & 0.15       &  9.55\\
S20      & {\em Gaia}-DR1 5631352971516952064 & 141.07585    & --31.70491     & 0.24       & $G = 20.26$\\
P194     & SDSS J095839.34+690912.1   & 149.66393    &   69.15337     & 2.15       & 15.47\\
S27      & USNO-B1.0 0300-0199877     & 162.53468    & --59.98405     & 1.09       & 15.55\\
S37      & SIMBAD\footnote{No optical counterpart was identified for this source in \cite{lansbury17}.  However, a search of the SIMBAD database \citep{wenger00} at the {\em Chandra} position for S37 indicates an association with HD 109573B, and the position given in this table is that of HD 109573B.} & 189.00231    & --39.87103     & 0.56       & 11.8\\
S43      & 2MASS J13015871--6358089   & 195.49464    & --63.96916     & 0.57       & $H = 12.05$\\
P316     & USNO-B1.0 0557-0301166     & 204.12056    & --34.23311     & 4.25       & 18.63\\
P340     & SDSS J143637.56+584303.3   & 219.15651    &   58.71761     & 0.86       & 13.98\\
P376     & SDSS J165350.78+393821.9   & 253.46158    &   39.63944     & 2.67       & 17.61\\
P408     & USNO-B1.0 0828-0514124     & 276.51929    &  --7.13521     & 0.26       & 15.88\\
P497     & USNO-B1.0 0662-0895815     & 353.61099    & --23.73609     & 1.87       & 12.49\\ \hline
\end{tabular}
\end{center}
\end{minipage}
\end{table}

\begin{table}
\caption{SIMBAD Identifications\label{tab:list_simbad}}
\begin{minipage}{\linewidth}
\begin{center}
\begin{tabular}{ccllc} \hline \hline
Serendip & NuSTAR          & SIMBAD Identifiers            & Wavelengths        & Type of\\
ID       & Name            &                               & Detected           & Source\\
         &                 &                               & Previously         &       \\ \hline\hline
S1       & J001639+8139.8  & 11 including HD 1165          & optical, IR        & Star\\
P82      & J042538--5714.5 & 6 including RX J0425.6--5714  & X-ray, UV, optical & CV/polar\\
P98      & J051626--0012.2 & 7 including V1193 Ori         & optical, IR        & CV/nova\\
P127     & J070014+1416.8  & -- & -- & --\\
P144     & J073959--3147.8 & -- & -- & --\\
P146     & J075611--4133.9 & 3 including TYC 7654-3811-1       & optical, IR        & Star\\
S20      & J092418--3142.2 & -- & -- & --\\
P194     & J095838+6909.2  & 4 including 2XMM J095839.2+690910 & X-ray, optical, IR & --\\
S27      & J105008--5958.8  & -- & -- & --\\
S37      & J123559--3951.9  & 10 including HD 109573B & X-ray, optical, IR & Star\\
S43      & J130157--6358.1  & 8 including 2RXP J130159.6--635806 & X-ray & HMXB\\
P316     & J133628--3414.1  & -- & -- & --\\
P340     & J143636+5843.0  & 8 including TYC 3866-132-1 & X-ray, optical, IR & Star\\
P376     & J165351+3938.5  & -- & -- & --\\
P408     & J182604--0707.9  & -- & -- & --\\
P497     & J233426--2343.9  & 1RXS J233427.8--234419 & X-ray & --\\ \hline
\end{tabular}
\end{center}
\end{minipage}
\end{table}

\begin{table}
\caption{Optical Emission Lines Detected\label{tab:opt_lines}}
\begin{minipage}{\linewidth}
\begin{center}
\begin{tabular}{cccccc} \hline \hline
Serendip & Element&${\lambda_{\rm c}}$\footnote{Measured wavelength in \AA.}&EW\footnote{Equivalent width in \AA.}&FWHM\footnote{Full-width at half-maximum in \AA.  Note that these values are not corrected for the instrumental resolution and should be taken as upper limits on the line widths.}&${F_{\rm line}}$\footnote{Intrinsic line flux in units of ${\rm erg\,\,cm}^{-2}\,\,{\rm s}^{-1}$.}\\
ID & & & & &\\\hline
P144 & H\,I (5-2) & $4344.5\pm2.9$  &  --$4.9\pm0.4$  &  $8.7\pm0.8$  &   ($8.1\pm3.5$)$\times 10^{-16}$\\
     & H\,I (4-2) & $4864.7\pm2.5$  &  --$2.1\pm0.7$  &  $14.2\pm5.3$ &   ($10.1\pm3.3$)$\times 10^{-16}$\\    
     & H\,I (3-2) & $6565.1\pm1.1$  &  --$13.2\pm1.4$ &  $28.2\pm2.4$ &   ($48.4\pm5.1$)$\times 10^{-16}$\\
     & He\,I     & $6681.0\pm2.4$  &  --$1.1\pm0.4$  &  $20.1\pm4.4$ &   ($3.9\pm 1.0$)$\times 10^{-16}$\\ \hline
S27 & H\,I (4-2)  & $4861.5\pm2.3$  &  --$6.9\pm2.0$  &  $5.5\pm2.6$  &   ($3.5\pm0.9$)$\times 10^{-15}$\\
    & H\,I (3-2)  & $6563.5\pm0.5$  &  --$28.2\pm4.4$ &  $8.9\pm1.0$  &   ($49.8\pm3.8$)$\times 10^{-15}$\\    
    & H\,I (16-3) & $8507.1\pm2.8$  &  --$6.7\pm1.6$  &  $23.8\pm3.7$ &   ($23.3\pm4.0$)$\times 10^{-15}$\\  
    & H\,I (15-3) & $8544.9\pm1.5$  &  --$13.0\pm2.6$ &  $23.5\pm3.2$ &   ($46.2\pm3.8$)$\times 10^{-15}$\\  
    & H\,I (14-3) & $8598.8\pm1.7$  &  --$5.6\pm1.3$  &  $14.6\pm2.0$ &   ($20.0\pm2.2$)$\times 10^{-15}$\\  
    & H\,I (13-3) & $8665.8\pm2.6$  &  --$5.0\pm1.3$  &  $12.2\pm2.2$ &   ($17.8\pm2.1$)$\times 10^{-15}$\\  
    & H\,I (12-3) & $8749.7\pm1.5$  &  --$9.1\pm1.4$  &  $16.0\pm1.6$ &   ($32.7\pm3.7$)$\times 10^{-15}$\\  
    & H\,I (11-3) & $8863.6\pm1.7$  &  --$5.9\pm1.7$  &  $12.2\pm1.4$ &   ($21.2\pm3.2$)$\times 10^{-15}$\\ \hline
P408 & H\,I (3-2)  & $6560.4\pm0.4$  &  --$4.9\pm0.4$  &  $8.7\pm0.8$  &   ($13.1\pm1.1$)$\times 10^{-16}$\\ \hline
P497 & H\,I (3-2)\footnote{This line is only marginally detected.} & $6555.4\pm 6.6$ & --$1.6\pm 0.9$  &  $23\pm 16$ &   ($2.1\pm 1.1$)$\times 10^{-14}$\\ \hline
\end{tabular}
\end{center}
\end{minipage}
\end{table}

\begin{table}
\caption{Observations for X-ray Energy Spectra\label{tab:obs}}
\begin{minipage}{\linewidth}
\begin{center}
\begin{tabular}{ccccc} \hline \hline
Serendip & Satellite        & ObsIDs      & Date of       & Exposure\\
   ID    &                  &             & Observation   & (ks)    \\ \hline
P144     & {\em NuSTAR}     & 60061351002 & 2014 April 20 & 22\\
         & {\em XMM-Newton} & 0501210201  & 2007 May 25   & 22\\
         & {\em Swift}      & 00080686001 & 2014 April 21 & 1.9\\ \hline
S27      & {\em NuSTAR}     & 30001024002 & 2013 July 17  & 293\\
         & {\em NuSTAR}     & 30001024003 & 2013 July 17  &    \\
         & {\em NuSTAR}     & 30001024005 & 2013 July 19  &    \\
         & {\em NuSTAR}     & 30001024007 & 2013 July 25  &    \\
         & {\em XMM-Newton} & 0654870101  & 2011 August 6 & 77\\ 
         & {\em Swift}      & 00080044001 & 2013 July 19  & 9.4\\
         & {\em Swift}      & 00080044002 & 2013 July 21  & 8.0\\ \hline
P408     & {\em NuSTAR}     & 60160688002 & 2015 May 3    & 20\\
         & {\em XMM-Newton} & 0650591501  & 2011 March 7  & 23\\
         & {\em Swift}      & 00081220001 & 2015 May 3    & 6.3\\ \hline
P497     & {\em NuSTAR}     & 60160832002 & 2015 July 30  & 18\\
         & {\em XMM-Newton} & 0760990101  & 2015 May 15   & 19\\
         & {\em XMM-Newton} & 0760990201  & 2015 November 17 & 20\\ 
         & {\em Swift}      & 00081308002 & 2015 July 30  & 6.2\\\hline
\end{tabular}
\end{center}
\end{minipage}
\end{table}

\begin{table}
\caption{Parameters for Fits to the {\em XMM-Newton} plus {\em NuSTAR} Energy Spectra\label{tab:parameters}}
\begin{minipage}{\linewidth}
\begin{center}
\begin{tabular}{cccccc} \hline \hline
Parameter      &   Units/Description   &    P144      &     S27    &    P408   &    P497\\ \hline
\multicolumn{6}{c}{Absorbed power-law ({\ttfamily tbabs*pegpwrlw})}\\ \hline
$N_{\rm H}$     &   $10^{22}$\,cm$^{-2}$ & $<$0.46          & $3.1^{+2.3}_{-1.5}$ & $0.9^{+0.4}_{-0.3}$ & $0.17\pm 0.02$\\
$\Gamma$      &   Photon index        & $1.4^{+0.5}_{-0.4}$ & $1.7^{+0.6}_{-0.5}$ & $2.9^{+0.6}_{-0.5}$ & $2.68\pm 0.11$\\
2--10\,keV Flux\footnote{This is the normalization for the {\ttfamily pegpwrlw} model, which is an unabsorbed flux.} & erg\,cm$^{-2}$\,s$^{-1}$ & $(1.5\pm 0.5)\times 10^{-13}$ & $(5.8^{+1.9}_{-1.5})\times 10^{-14}$ & $(1.5^{+0.7}_{-0.6})\times 10^{-13}$ & $(4.9\pm 1.1)\times 10^{-13}$\\
8--24\,keV Flux        & erg\,cm$^{-2}$\,s$^{-1}$ & $(1.9^{+0.7}_{-1.0})\times 10^{-13}$ & $(5.4^{+2.0}_{-2.7})\times 10^{-14}$ & $(3.7^{+1.9}_{-2.4})\times 10^{-14}$ & $(1.5\pm 0.4)\times 10^{-13}$\\
$C_{XMM}/C_{NuSTAR}$ & -- & $0.14^{+0.15}_{-0.08}$ & $0.76^{+0.35}_{-0.24}$ & $0.14^{+0.13}_{-0.07}$ & $0.26^{+0.08}_{-0.06}$\\
$C_{Swift}/C_{NuSTAR}$\footnote{The fits that resulted in the parameters and $\chi^{2}$ values in this table did not include the {\em Swift}/XRT data.  We performed a second round of fits with the XRT data to determine $C_{Swift}/C_{NuSTAR}$.} & -- & $1.2^{+1.8}_{-0.8}$ & $<$0.77 & $0.6^{+0.6}_{-0.3}$ & $0.9^{+0.3}_{-0.2}$\\
$\chi^{2}/\nu$  & -- & 8.1/12 & 27.2/15 & 14.5/13 & 91/83\\ \hline
\multicolumn{6}{c}{Absorbed thermal bremsstrahlung ({\ttfamily tbabs*bremss})}\\ \hline
$N_{\rm H}$     &   $10^{22}$\,cm$^{-2}$  &  $<$0.32        & $2.4^{+1.7}_{-1.0}$ &  $0.59^{+0.26}_{-0.18}$ & $0.048\pm 0.013$\\
$kT$           &  keV                  & $>$9            & $>$9             & $2.3^{+1.2}_{-0.8}$      & $1.61^{+0.17}_{-0.15}$\\
Normalization\footnote{$\frac{3.02\times 10^{-15}}{4\pi\,D^{2}}\int{n_{e}\,n_{I}\,dV}$, where $D$ is the distance to the source (in cm), $n_{e}$ and $n_{I}$ are the electron and ion densities (in cm$^{-3}$), and $V$ is the volume of the emitting region (in cm$^{3}$).} &  -- & $(3.9^{+1.5}_{-1.1})\times 10^{-5}$ & $(1.4^{+0.5}_{-0.3})\times 10^{-5}$ & $(1.4^{+2.2}_{-0.8})\times 10^{-4}$ & $(9\pm 3)\times 10^{-4}$\\
$C_{XMM}/C_{NuSTAR}$ & -- & $0.14^{+0.11}_{-0.07}$ & $0.78^{+0.34}_{-0.24}$ & $0.14^{+0.11}_{-0.06}$ & $0.18^{+0.06}_{-0.04}$\\
$\chi^{2}/\nu$  & -- & 7.5/12 & 26.7/15 & 12.3/13 & 109/83\\ \hline
\end{tabular}
\end{center}
\end{minipage}
\end{table}

\begin{table}
\caption{Source Classifications\label{tab:classifications}}
\begin{minipage}{\linewidth}
\begin{center}
\begin{tabular}{cclll} \hline \hline
Serendip & NuSTAR          & Classification         & Other possible          & Primary method\\
ID       & Name            &                        & classifications\footnote{WD indicates the possibility of a white dwarf binary companion.  The Star+WD systems could also be called CVs.  As described in the text of the paper, AB is an active binary, consisting of two stars with at least one producing coronal X-ray emission.}        & of classification\\ \hline\hline
S1       & J001639+8139.8  & Star(K0)               & --                      & SIMBAD\\
P82      & J042538--5714.5 & CV/polar               & --                      & SIMBAD\\
P98      & J051626--0012.2 & CV/nova                & --                      & SIMBAD\\
P127     & J070014+1416.8  & Star(F)                & Star(F)+WD or AB        & Optical spectrum\\
P144     & J073959--3147.8 & Black hole LMXB        & CV                      & Optical and X-ray spectra\\
P146     & J075611--4133.9 & Star                   & Star+WD                 & SIMBAD and X-ray flux\\
S20      & J092418--3142.2 & LMXB                   & CV                      & X-ray flux and optical spectrum\\
P194     & J095838+6909.2  & Star(M)                & Star(M)+WD or AB        & Optical spectrum\\
S27      & J105008--5958.8 & HMXB                   & Star(Be)+WD or Star(Be) & Optical and X-ray spectra\\
S37      & J123559--3951.9 & Star(M2.5)             & --                      & SIMBAD\\
S43      & J130157--6358.1 & HMXB                   & --                      & SIMBAD\\
P316     & J133628--3414.1 & Star(K-G)              & Star(K-G)+WD or AB      & Optical spectrum\\
P340     & J143636+5843.0  & Star                   & --                      & SIMBAD\\
P376     & J165351+3938.5  & Star(K-G)              & Star(K-G)+WD or AB      & Optical spectrum\\
P408     & J182604--0707.9 & CV                     & X-ray binary            & Optical and X-ray spectra\\
P497     & J233426--2343.9 & CV                     & AB                      & Optical and X-ray spectra\\ \hline
\end{tabular}
\end{center}
\end{minipage}
\end{table}

\end{document}